\documentclass[11pt]{article}

\usepackage[english]{babel}
\usepackage[
    a4paper,
    left=3cm,
    top=3.5cm,
    right=3cm,
    bottom=3.5cm
]{geometry}

\usepackage[T1]{fontenc}
\usepackage[utf8]{inputenc}
\usepackage{graphicx}
\usepackage[hidelinks]{hyperref}
\usepackage{mathtools, amsmath, amssymb, bm}
\usepackage[scr=boondoxo]{mathalpha}
\usepackage{cleveref}
\usepackage[final,nopatch=footnote]{microtype}
\usepackage[export]{adjustbox}
\usepackage[title]{appendix}
\usepackage{color, xcolor}

\usepackage[authoryear,longnamesfirst,round]{natbib}

\usepackage{algpseudocode, algorithm}
\algrenewcommand\algorithmicrequire{\textbf{Input:}}
\algrenewcommand\algorithmicensure{\textbf{Output:}}

\DeclareMathOperator*{\argmax}{arg\,max}
\DeclareMathOperator*{\argmin}{arg\,min}
\DeclareMathOperator*{\Vol}{Vol}
\newcommand{\PrU}{\operatorname{\mathrm{Pr}}_{\bm{U}}}
\newcommand{\set}[1]{\left\{#1\right\}}

\title{\bf Active learning strategy for excursion-set confidence regions of functional simulator outputs}
\date{}

\usepackage{authblk}
\author[1,2]{Lucas Brunel}
\author[1]{Mathieu Balesdent}
\author[1]{Lo\"ic Brevault}
\author[2]{Rodolphe Le~Riche}
\author[3]{Bruno Sudret}
\affil[1]{ONERA, Universit\'e Paris-Saclay, Palaiseau, France}
\affil[2]{LIMOS (CNRS, Mines Saint-\'Etienne and Universit\'e Clermont Auvergne), France}
\affil[3]{ETH Zurich, Chair of Risk, Safety and Uncertainty Quantification, Switzerland}

\begin{document}

\maketitle

\begin{abstract}
    Estimating excursion set confidence regions seeks to identify regions where a function may exceed some threshold with a given confidence level.
    This paper focuses on estimating such confidence regions in cases where the function has random inputs and a functional output that is returned all at once.
    We develop a surrogate-based approach for estimating the confidence region, combining principal component analysis and Gaussian process regression.
    An active learning strategy is also introduced, based on a max-min criterion that selects new samples which are likely to reduce the uncertainty in the confidence region. This strategy leverages efficient sampling of the Gaussian process through a Karhunen-Lo\`eve expansion.

    The proposed approach is applied to estimate the confidence regions of three case studies: a synthetic function, the surface pressure coefficient distribution of a hypersonic vehicle, and the glide-back trajectory of a reusable launcher first stage.
    The method demonstrates efficiency in accurately estimating the confidence region while reducing sources of modeling uncertainties.
    It is benchmarked against reference methods from the literature.
    Relevant metrics for assessing the confidence region estimation performance are discussed.
\end{abstract}

\section{Introduction}

In many scientific and engineering disciplines, identifying the regions in a system input space that lead to undesirable outcomes is critical.
These regions, where a function of interest exceeds or falls below a specified threshold, are commonly referred to as \emph{excursion sets} \citep{molchanovTheory2005}.
Practical applications span a wide range of domains, including pollution control in the automotive industry \citep{elamriFeasible2023}, chemical measurements in the environment \citep{frenchCredible2016}, and marine science \citep{fossumLearning2021}.
When the function of interest is random, the corresponding excursion set is a random set.
Random set theory \citep{molchanovTheory2005} provides a framework to describe and quantify the uncertainty associated with such sets.
Throughout this paper, random excursion sets are simply referred to as excursion sets.

A number of statistical approaches have been proposed to analyze excursion sets.
Among them, \citet{chevalierEstimating2013} use the Vorob'ev expectation and deviation, which were applied in nuclear safety, to determine the safe configurations of a nuclear reactor \citep{chevalierFast2013}.
Another key concept, which is the focus of this research, is the notion of \emph{confidence regions} for excursion sets \citep{frenchSpatiotemporal2013,bolinExcursion2015}.
A confidence region is a set that, with a given probability $\alpha$, either contains or is contained within the excursion set.
When the confidence is about containing the excursion set, the confidence region is referred to as an outer confidence set in \citet{bulthuisMultifidelity2020}.
Conversely, when it is about being contained in the excursion set, it is called a conservative estimate in \citet{azzimontiAdaptive2021} and an inner confidence set in \citet{bulthuisMultifidelity2020}.
Confidence regions provide an estimate of where the excursion is likely to occur, with a specified probability, accounting for the uncertainty of the underlying function producing the excursion.
This concept has proven valuable in a variety of domains, including air pollution monitoring \citep{bolinExcursion2015}, glaciology \citep{bulthuisMultifidelity2020}, and nuclear safety \citep{azzimontiAdaptive2021}.

The function that defines the excursion set often represents a physical system modeled by a computationally expensive simulator.
In such cases, estimating a confidence region can be prohibitively costly.
A widely adopted solution is to use an approximation of the simulator, known as a \emph{surrogate model}, which is much cheaper to evaluate, albeit at the cost of introducing some approximation error \citep{khatouriMetamodeling2022}.
To train the surrogate model, the simulator must be evaluated at a set of input configurations, known as a \emph{design of experiments} (DoE).

There are two main strategies to construct a DoE: space-filling methods and active learning, also known as adaptive sampling.
Space-filling methods \citep{santnerSpacefilling2018} aim to distribute the samples across the input space to gather as much information as possible from the corresponding output samples.
In contrast, active learning begins with a small-sized initial DoE, typically a space-filling design, and iteratively adds new samples based on a criterion that quantifies the informativeness of candidate samples.
This allows the analyst to make the most out of a limited computational budget, while controlling the approximation error.
The criterion depends on the intended use of the surrogate.
In this work, we focus on active learning designed to estimate confidence regions of excursion sets, resulting from uncertainty propagation through a costly simulator with functional outputs.

Several active learning strategies have been proposed in related contexts. \citet{azzimontiAdaptive2021} developed a stepwise uncertainty reduction strategy where the simulator with a scalar-valued output is replaced by a Kriging model \citep{rasmussenGaussian2005}.
They proposed multiple selection criteria for sequentially choosing new samples, and used their method to compute a conservative estimate of the excursion set.
This approach is not readily applicable to functional outputs.

Considering a simulator with a functional output, \citet{bulthuisMultifidelity2020} proposed an approach combining a polynomial chaos expansion \citep[PCE;][]{ghanemSpectral1991,xiuWienerAskey2002,soizePhysical2004} to model an intermediate quantity related to the confidence region, using kernel density estimation \citep[KDE;][]{parzenEstimation1962,wandKernel1994} to approximate the probability of exceeding the threshold at each location in the mesh.
Their active learning criterion is designed to reduce the PCE modeling error but it does not improve the KDE approximation of the pointwise probability, thus limiting the overall accuracy of the estimated confidence region.
Furthermore, the reliance on KDE can be problematic to model the tails of the distributions, which are regions of prime importance in reliability analysis.
Notwithstanding the pointwise application of the KDE, the use of random sets ensures that the joint distribution of the underlying function is accounted for.
In summary, available methods either are not readily applicable to functional outputs, or remain costly and do not fully take advantage of available data. The current study aims to tackle these limitations.

This study focuses on deterministic simulators with functional outputs that return results over an entire domain (spatial, temporal, etc.) in a single evaluation.
The uncertainty stems from the random inputs, propagated through the simulator.
Let $y(\bm{U}, \bm{x})\in\mathbb R$ be the simulator output at location $\bm{x}\in\mathbb{X}\subseteq\mathbb R^{d_{\bm x}}$, given the uncertain input vector $\bm{U}\in\mathbb{U}\subset\mathbb R^{d_{\bm u}}$, assuming its probability density function $f_{\bm U}$ is given.
The collection of all mesh locations is denoted $\mathcal{X}=\{\bm{x}_1,\dots,\bm{x}_{n_{\bm{x}}}\}$, with $n_{\bm x}$ being the number of mesh nodes.
A single run of the computationally expensive simulator at an input $\bm{u}_i$ yields the full vector of outputs $\bm{y}(\bm u_i)=[y(\bm{u}_i,\bm x_1),\dots,y(\bm{u}_i,\bm x_{n_{\bm x}})]^\top$.
In this work, we assume that the mesh does not change with the value of the input vector.
Otherwise, it might require an additional step of projection onto a common mesh. 
The objective is to estimate a confidence region for the excursion set associated with this functional output.
To do so, we propose a dedicated active learning strategy that maximizes accuracy within a limited computational budget.

The structure of this paper is as follows.
\Cref{sec:random-sets} introduces the computation of the confidence regions for excursion sets.
\Cref{sec:surrogate-modeling} presents the surrogate model for functional outputs used throughout the study.
\Cref{sec:methodology} details the surrogate-based estimation of the confidence region and the associated active learning strategy.
\Cref{sec:numerical-experiments} applies the proposed methodology to three case studies: a challenging synthetic example, the surface pressure distribution of a hypersonic vehicle, and the return-to-launch-site trajectory of a reusable launcher first stage.
Finally, \Cref{sec:conclusion} summarizes the method and outlines potential directions for future works.

\section{Confidence regions for excursion sets in uncertainty propagation for simulators with functional outputs}\label{sec:random-sets}

An \emph{excursion set}, denoted by $\mathcal{E}(\bm U)$, is the set of all locations $\bm x$ in the mesh $\mathcal X$ where the simulator output falls within a target range of values $\mathbb T=[t,+\infty)$, $\mathbb T=(-\infty,t]$ or $\mathbb{T}=[t_\mathrm{low},t_\mathrm{high}]$:
\begin{equation}\label{eq:excursion}
    \mathcal{E}(\bm U)=\set{\bm x\in\mathcal X:y(\bm U, \bm x)\in \mathbb T},
\end{equation}

Considering the uncertainty propagation context, the simulator output is modeled as a random function.
Therefore, $\mathcal{E}(\bm U)$ is a random set \citep{molchanovTheory2005}.
Based on this, the \emph{coverage probability function} $p(\bm x)$ is defined by
\begin{equation}
    p(\bm x) = \PrU[\bm x\in\mathcal{E}(\bm U)]=\PrU[y(\bm U,\bm x)\in\mathbb T],
\end{equation}
where $\PrU[\cdot]$ denotes the probability of the event defined in between brackets under the probability measure $d\PrU=f_{\bm{U}}(\bm{u})d\bm{u}$.

Among the several metrics to characterize random sets, the Vorob'ev $\rho$-quantile \citep{molchanovTheory2005} is defined as the set of mesh locations where the coverage probability function is greater than or equal to $\rho$, that is,
\begin{equation}\label{eq:vorobev-quantile}
    \mathcal{Q}_\rho=\{\bm x\in\mathcal X:p(\bm x) \geq \rho\}.
\end{equation}

Alternatives to the Vorob'ev quantile include the Vorob'ev expectation and deviation \citep{molchanovTheory2005}.
Another important metric is a \emph{confidence region}, which is a closed subset $\mathcal R$ of $\mathcal X$, such that the probability that it contains the excursion set is greater than or equal to a prescribed $\alpha$:
\begin{equation}\label{eq:proba-confidence-region}
    \PrU[\mathcal E(\bm U)\subset\mathcal R] \geq \alpha.
\end{equation}

For a given confidence level $\alpha$, the region $\mathcal{R}$ is not unique in general as any of its superset which is a subset of $\mathcal{X}$ also satisfies \Cref{eq:proba-confidence-region}.
The most interesting of such regions is the smallest, so as to avoid over-conservativeness.
The region $\mathcal{R}$ that minimizes its volume is denoted by $\mathcal{C}_\alpha$, and defined by
\begin{equation}
    \mathcal{C}_{\alpha} = \argmin_{\mathcal R\in\mathfrak R}\{\Vol(\mathcal R):\PrU[\mathcal E(\bm U)\subset\mathcal R] \geq \alpha\},
\end{equation}
where $\mathfrak R$ is a family of closed subsets of $\mathcal X$ and $\Vol(\cdot)$ denotes the volume. Depending on $\dim(\mathbb X)$, the latter might be a length, an area, a classical volume or a hyper-volume.
In the following, we simply use the term \emph{volume}, regardless of the dimensionality of $\mathbb X$.

Following prior works \citep{frenchSpatiotemporal2013, bolinExcursion2015, azzimontiAdaptive2021, bulthuisMultifidelity2020}, $\mathcal R$ is sought within the parametric family of Vorob'ev $\rho$-quantiles, parametrized by $\rho\in[0,1]$, and denoted by $\mathscr{V}$.
Alternative families may be used for $\mathcal R$ \citep{frenchSpatiotemporal2013,bolinExcursion2015,frenchCredible2016,bulthuisMultifidelity2020}.
Some involve multiple parameters, but do not admit the simplification that is discussed below and which is needed for the proposed active learning strategy described in \Cref{sec:methodology}.
For $\mathcal{R}\in\mathscr{V}$, minimizing the volume of $\mathcal{R}$ boils down to maximizing $\rho$ \citep{frenchSpatiotemporal2013}.

In order to transform the confidence-set estimation process into a simpler quantile estimation, the problem needs to be reformulated \citep{frenchSpatiotemporal2013}.
First, the condition in \Cref{eq:proba-confidence-region} is rewritten.
If $\mathcal{E}(\bm{u})\neq\emptyset$, the event $\mathcal{E}(\bm{u})\subset\mathcal{R}$ means that every $\bm{x}\in\mathcal{E}(\bm{u})$ satisfies $p(\bm{x})\geq\rho$, according to the definition of the Vorob'ev quantile in \Cref{eq:vorobev-quantile}.
This condition is met if the minimum coverage probability over the excursion set is at least $\rho$.
To ease further calculations, let us introduce the auxiliary random variable $\chi(\bm{U})=\min\{p(\bm{x}):\bm{x}\in\mathcal{E}(\bm{U})\}$ \citep{bulthuisMultifidelity2020}.
\Cref{eq:proba-confidence-region} then becomes
\begin{equation}\label{eq:simplification}
\begin{aligned}
    \PrU[\mathcal E(\bm U)\subset\mathcal R] & = \PrU[\min\{p(\bm x):\bm x\in\mathcal E(\bm U)\}\geq\rho] {}& \\
    & = \PrU[\chi(\bm U) \geq \rho] {}& \\
    & = 1 - \PrU[\chi(\bm U) < \rho] {}& \;\;\geq \alpha.
\end{aligned}
\end{equation}

If, for a given realization $\bm{u}$, $\mathcal{E}(\bm{U})=\emptyset$, then $\chi(\bm{u})$ is not defined as currently formulated.
To circumvent this issue, authors often assume non-empty excursion sets \citep{frenchSpatiotemporal2013,bulthuisMultifidelity2020}.
Otherwise, \citet{frenchCredible2016} define $\chi(\bm{u})=0$ when $\mathcal{E}(\bm{u})=\emptyset$ (see their Algorithm~1).
Besides, the map $\bm{u}\mapsto\chi(\bm{u})$ may be discontinuous.

Finally, the optimal region is $\mathcal C_{\alpha}=\mathcal Q_{\rho^*}$, where the optimal threshold $\rho^*$ is obtained by solving the quantile problem:
\begin{equation}
    \rho^*=\max \set{\rho\in[0,1] :\PrU[\chi(\bm U) < \rho] \leq 1-\alpha},
\end{equation}
that is, $\rho^*$ is the $(1-\alpha)$-quantile of the auxiliary variable $\chi(\bm U)$.
For brevity, the region $\mathcal C_{\alpha}$ is referred to as \emph{the confidence region} in the remainder of this work.
Note that thanks to the above rewriting, the estimation of the confidence region amounts to estimating a quantile of the scalar random variable $\chi(\bm{U})$, thereby reducing the computational complexity.

In the sequel, this quantile and all intermediate quantities are estimated using Monte Carlo simulation (MCS).
For the estimates to be accurate when $\alpha$ is small, the sample size needs to be large.
Because of the significant computational cost of a complex system simulator, using it with MCS is often intractable.
To alleviate this problem, a classical approach consists in replacing the costly simulator with a surrogate model.

\section{Surrogate modeling with functional outputs}\label{sec:surrogate-modeling}

Building and training a surrogate with a functional output is challenging.
To lower the complexity of this process, dimensionality reduction \citep{vandermaatenDimensionality2009,leeNonlinear2007} can be employed to represent high-dimensional data using a significantly smaller number of variables.
Although this inevitably leads to some loss of information, it enables the surrogate model to map inputs not directly to the high-dimensional output space, but rather to a low-dimensional \emph{latent space}.
A variety of approaches are available to construct such mappings to the latent space \citep{khatouriMetamodeling2022}, that we call the \emph{latent surrogate}.

\subsection{Dimensionality reduction with principal component analysis}\label{sec:pca}

\emph{Principal component analysis} (PCA) is a linear dimensionality reduction technique \citep{jolliffePrincipal2002}.
In the model order reduction community, it is often referred to as \emph{proper orthogonal decomposition} (POD), and in the statistics community, \emph{Karhunen-Lo\`eve expansion} \citep{liangProper2002}.
It projects the $n_{\bm x}$-dimensional outputs $\bm y(\bm U)$ onto the vector space spanned by the eigenvectors of the empirical covariance matrix computed from a set of samples called \emph{snapshots}.

Suppose that $n$ samples $\{\bm u_1,\dots,\bm u_n\}$ of the input vector have been generated, and the simulator has been run to get the corresponding samples of the output $\{\bm y(\bm u_1),\dots,\bm y(\bm u_n)\}$.
Note that the samples of $\bm u$ do not need to be drawn from the distribution of $\bm U$.
PCA begins by subtracting the mean $\bar{\bm y}=\frac{1}{n}\sum^n_{i=1}\bm y(\bm u_i)$ from all snapshots.
The $n$-by-$n_{\bm x}$ matrix of the centered snapshots is denoted by $\bar{\mathbf{Y}}=[\bm y(\bm u_1) - \bar{\bm y}, \dots,\bm y(\bm u_n) - \bar{\bm y}]^\top$.
Then, the sample, or empirical, covariance matrix is computed as $\mathbf{C}=\frac{1}{n-1}\bar{\mathbf{Y}}^\top\bar{\mathbf{Y}}$.
The eigendecomposition of $\mathbf{C}$ gives the eigenvalues $\lambda_j\in\mathbb{R}$ and the eigenvectors $\bm v_j\in\mathbb{R}^{n_{\bm x}}$, with $j\in\{1,\dots,n\}$.
Usually, instead of an eigendecomposition of $\mathbf{C}$, a singular value decomposition \citep{eckartApproximation1936} of $\bar{\mathbf{Y}}$ is performed.

To actually reduce the dimensionality of the output space, one needs to select a subset of these eigenvectors.
A popular approach consists of using a \emph{relative information content} (RIC) threshold \citep{pinnauModel2008}, defined by
\begin{equation}
    \mathrm{RIC}(r) = \frac{\sum^r_{j=1}\lambda_j}{\sum_{j=1}^{n}\lambda_j}.
\end{equation}

Given a threshold $q\in[0,1]$, typically $q=0.95$ or 0.99, the number of retained components is defined by

\begin{equation}
    d_{\bm z}=\argmin_{r \in \{1,\ldots,n_{\bm x}\}}\{\mathrm{RIC}(r):\mathrm{RIC}(r)\geq q, q\in[0,1]\},
\end{equation}
where $d_{\bm{z}}$ is the dimensionality of the latent space.
This results in a subset of $d_{\bm{z}}$ eigenvectors collected in the projection matrix $\mathbf{V}=[\bm v_1,\dots,\bm v_{d_{\bm z}}]^\top$.

To obtain the value of the latent variables $\bm z(\bm u)$ corresponding to a sample $\bm y(\bm u)$, the latter is projected onto the latent space using $\mathbf{V}$, that is, $\bm z(\bm u)=\mathbf{V}^\top(\bm y(\bm u) - \bar{\bm y})$.
Conversely, a high-dimensional field corresponding to a latent variable vector $\bm z(\bm u)$ is computed as $\bm y(\bm u)=\mathbf V\bm z(\bm u) + \bar{\bm y}$.
As a consequence, to make predictions of the simulator output $\bm y(\bm u)$, we need to first predict the latent vector $\bm z(\bm u)$.
To do this, we need to map the input space to the latent space.

\subsection{Kriging for latent surrogate modeling}\label{sec:kriging}

To map the input space to the latent space, the latent variables are modeled using \emph{Kriging} \citep{rasmussenGaussian2005}, also called \emph{Gaussian process regression}.
The main reason for this choice is that it provides an associated predictive uncertainty along with a prediction, which can be used in the subsequent active learning process.
A Gaussian process (GP) is a random function of which any finite collection of random variables follows a joint multivariate normal distribution.
Kriging with a scalar-valued output approximates a function $z$ by a GP $Z:\mathbb U\rightarrow\mathbb R$ conditioned on a set of observations $(\bm u_i,z(\bm u_i))$, $i\in\{1,\dots,n\}$.

A GP is fully described by its mean and covariance functions, respectively $\mu$ and $k_{\bm\theta}$, with $\bm\theta$ a set of hyperparameters.
$\mu$ and $k_{\bm\theta}$ represent the prior knowledge on the GP in the Bayesian framework.
In the absence of knowledge on the trend of the function $z$, one can assume an unknown constant mean function \citep{rasmussenGaussian2005}.
Such a model is called ordinary Kriging.
There is a variety of choices for the covariance function, also called the kernel, which incorporates assumptions on the modeled function, such as the periodicity and differentiability. Examples include the squared-exponential and the Mat\'ern family of kernels \citep[see][for more examples]{santnerSpacefilling2018}.
To determine the value of the hyperparameters that should be used to make predictions, a classical method is to maximize the marginal likelihood \citep{rasmussenGaussian2005}.

Let $\mathscr{U}=\{\bm u_1,\dots,\bm u_n\}$ and $\mathscr{Z}=\{z(\bm u_1),\dots,z(\bm u_n)\}$ be the observations of the input and output samples.
At an unobserved $\bm u\in\mathbb U$, the GP conditioned on the observations $Z_n(\bm{u})=Z(\bm{u})|(\mathscr{U,Z})$ has a posterior distribution of mean $m(\bm{u})$ and variance $s^2(\bm{u})$:
\begin{align}
    & m(\bm u) = \mu + \bm{k}^\top(\mathbf{K} + \sigma^2\mathbf{I})^{-1}(\bm z_\mathrm{obs} - \bm\mu), \\
    & s^2(\bm u) = k_{\bm\theta}(\bm u,\bm u) - \bm{k}^\top(\mathbf{K} + \sigma^2\mathbf{I})^{-1}\bm{k},
\end{align}
where $\bm{k} = [k_{\bm\theta}(\bm u, \bm u_1),\dots,k_{\bm\theta}(\bm u, \bm u_n)]^\top$, $\mathrm{K}_{i,j} = k_{\bm\theta}(\bm u_i, \bm u_j)$ for $i,j\in\{0,1,\dots,n\}^2$, $\bm z_\mathrm{obs}=[z(\bm u_1), \dots, z(\bm u_n)]^\top$, $\bm\mu$ is a $n$-sized vector where all entries are $\mu$, $\mathbf{I}$ is the $n$-by-$n$ identity matrix, and $\sigma^2$ is a Gaussian homoscedastic noise variance parameter, modeling possible observational noise.

In our approach, one Kriging model is built per latent variable, leading to $d_{\bm z}$ GPs with posterior mean $m_i$, $i\in\{1,\dots,d_{\bm z}\}$.
They enable the generation of corresponding predictions of the simulator output by combining them with PCA:
\begin{equation}
    \hat{\bm{y}}(\bm u) = \mathbf{V}\hat{\bm{z}}(\bm u) + \bar{\bm y},
\end{equation}
where $\hat{\bm{z}}(\bm u)=[m_1(\bm u),\dots,m_{d_{\bm z}}(\bm u)]^\top$.

The accuracy of this prediction depends on several factors, including both the PCA and the training of the Kriging model.
As is explained in the introduction, active learning then allows us to strategically generate the training samples to improve the surrogate estimation of the quantity of interest, which is the confidence regions $\mathcal{C}_{\alpha}$ in this work.

\section{Active learning strategy for excursion-set confidence regions}\label{sec:methodology}

This section begins by describing the process of estimating the confidence region using the surrogate model introduced earlier.
Next it outlines how the uncertainty associated with this estimate is quantified.
Finally, it presents the active learning strategy developed to efficiently refine the confidence region by leveraging the uncertainty in the estimate.

\subsection{Surrogate-based estimation of a confidence region}\label{sec:surrogate-estimation}

Uncertainty propagation is conducted using crude Monte Carlo simulation.
The Monte Carlo input samples are collected in $\mathscr{U}_\mathrm{MCS}=\{\bm{u}_1,\dots,\bm{u}_{n_\mathrm{MCS}}\}$, and drawn from the distribution of $\bm{U}$.
For each sample, the surrogate provides a corresponding prediction of the functional output $\hat y(\bm u_i,\mathcal X)$ over the entire mesh.
These predicted outputs are used to estimate the coverage probability function with
\begin{equation}\label{eq:coverage}
    \hat{p}(\bm x)=\frac{1}{n_\mathrm{MCS}}\sum^{n_\mathrm{MCS}}_{i=1}\mathbf{1}_\mathbb T(\hat y(\bm u_i,\bm x)),
\end{equation}  
where $\mathbf{1}_\mathbb T(\hat y(\bm u_i,\bm x))$ is the indicator function which takes the value 1 if $\hat y(\bm u_i,\bm x)\in\mathbb T$, and 0 otherwise.

The surrogate outputs are also used to determine the corresponding predicted excursion sets, defined by $\hat{\mathcal E}(\bm u_i)=\{\bm x\in\mathcal X:\hat y(\bm u_i,\bm x)\in\mathbb T\}$.
To be able to work with empty excursion sets with the simplification introduced in \Cref{sec:random-sets}, we propose an alternative definition for $\chi(\bm{U})$, denoted by $\hat{\chi}$:
\begin{equation}\label{eq:def-chi}
\begin{split}
    \hat{\chi}:{} & \mathbb{U}\rightarrow[0,1] \\
    & \bm{u}\mapsto\left\{
        \begin{matrix*}[l]
            \min\{\hat p(\bm x):\bm x\in\hat{\mathcal E}(\bm u)\} & \text{if $\hat{\mathcal{E}}(\bm{u})\neq\emptyset$},\\
            1 & \text{otherwise.}
        \end{matrix*}
    \right.
\end{split}
\end{equation}
The rationale for this definition is provided in \Cref{app:proof}.

For each excursion set, the auxiliary variable $\hat{\chi}$ is computed with \Cref{eq:def-chi}.
The optimal parameter $\hat{\rho}^*$ is the empirical $(1-\alpha)$-quantile, obtained by first sorting the values in the Monte Carlo sample $\{\hat{\chi}(\bm u_1),\dots,\hat{\chi}(\bm u_{n_\text{MCS}})\}$ such that the $\hat{\chi}(\bm{u})$ are in ascending order and then taking:
\begin{equation}\label{eq:quantile}
    \hat{\rho}^* = \hat{\chi}(\bm u_l), \hspace{2em}\text{with}\hspace{2em}l=\lceil(1 - \alpha) n_\mathrm{MCS}\rceil,
\end{equation}
where $\lceil\cdot\rceil$ is the ceiling function.
Finally, the predicted confidence region is $\hat{\mathcal{C}}_{\alpha}=\{\bm x\in\mathcal X:\hat{p}(\bm x)\geq \hat{\rho}^*\}$.

This modeling process leads to various sources of uncertainties: the Monte Carlo approximation for probability estimation, the dimensionality reduction error introduced by PCA, the predictive uncertainty of Kriging, the statistical uncertainty when drawing the initial DoE, and the parametrization of the confidence region.
Among these, the active learning process should mitigate the uncertainty associated with the selection of training samples.
In our active learning criterion, we focus on the modeling uncertainty stemming from the Kriging model, while the uncertainty associated with the choice of the initial DoE is characterized empirically in the experiments.
The uncertainties associated with the Monte Carlo sample size and the parameterization of the shape of the confidence region (through the Vorob'ev quantile) are left for future research.

\subsection{Uncertainty on the estimation of a confidence region}\label{sec:uncertainty-C}

In this work, we focus on the uncertainty caused by the intrinsic variability of the latent GPs.
While propagating the analytical predictive uncertainty of the GPs through the PCA is straightforward, the challenge lies in the subsequent propagation steps. Indeed, to the best of the authors' knowledge, there is currently no established analytical method to propagate GP uncertainty through the min operator over the random excursion set, which defines $\hat{\chi}(\bm{u})$. Consequently, we adopt a sampling strategy.

For this purpose, we generate $n_\mathrm{rea}$ realizations of the GPs for each sample $\bm{u}\in\mathscr{U}_\mathrm{MCS}$.
Rather than sampling the conditioned GP directly, we first sample an unconditioned GP and then apply a correction based on the training data, following the approach in \citet{chilesGeostatistics2012} and \citet{legratietBayesian2014}.
This formulation enables the use of a Karhunen-Lo\`eve expansion \citep{sudretStochastic2000} of the unconditioned GP, approximated using the Nystr\"om method \citep{betzNumerical2014}.
It significantly reduces the computational cost when the GP is to be sampled for a large number of input points, which is the case here due to the Monte Carlo simulation.

The $j$-th realization of the GPs prediction at the $i$-th Monte Carlo sample is denoted by $\hat{\bm z}_{\mathrm{GP}_j}(\bm u_i)$, where $j\in\{1,\dots,n_\mathrm{rea}\}$ and $i\in\{1,\dots,n_\mathrm{MCS}\}$.
These realizations are then mapped back to the original output space using the PCA basis:
\begin{equation}
    \hat{\bm y}_{\mathrm{GP}_j}(\bm u_i)=\mathbf{V}\hat{\bm z}_{\mathrm{GP}_j}(\bm u_i)+\bar{\bm y}.
\end{equation}

This yields $n_\mathrm{rea}$ realizations of the functional output for a given $\bm{u}_i$.
Using the methodology described in \Cref{sec:surrogate-estimation}, these outputs enable the generation of $n_\mathrm{rea}$ corresponding realizations of the following quantities: the excursion set $\hat{\mathcal{E}}_{\mathrm{GP}_j}(\bm u_i)$, the coverage probability function $\hat{p}_{\mathrm{GP}_j}$, the auxiliary variable $\hat{\chi}_{\mathrm{GP}_j}(\bm u_i)$, and the estimated threshold $\hat{\rho}^*_{\mathrm{GP}_j}$.

These multiple realizations provide a way to quantify the uncertainty in the estimation of the confidence region linked to the Kriging uncertainty.
In particular, the collection of $\hat{\rho}^*_{\mathrm{GP}_j}$ values is used in the next section to define an active learning criterion that targets areas of highest uncertainty.
Specifically, we use the 0.1- and 0.9-quantiles of $\hat{\rho}^*$, obtained by first sorting the values of $\hat{\rho}^*$ in ascending order and then taking:
\begin{equation}
    q_{\hat{\rho}^*,\beta} = \hat{\rho}^*_{\mathrm{GP}_j}, \hspace{2em}\mathrm{with}\hspace{2em}j=\lceil\beta\times n_\mathrm{rea}\rceil,
\end{equation}
where $\beta\in\{0.1,0.9\}$.
The next section discusses the choice of these values.

\subsection{Active learning of a confidence region}

The proposed methodology begins with an initial DoE $\mathscr U=\{\bm u_1, \dots, \bm u_n\}$, generated using a latin hypercube sampling (LHS) scheme \citep{santnerSpacefilling2018}.
The corresponding values of the simulator output are $\mathscr Y=\{\bm y(\bm u_1), \dots, \bm y(\bm u_n)\}$.
New samples are then added sequentially, according to the following active learning criterion, referred to as \emph{max-min} \citep{johnsonMinimax1990,lacazeGeneralized2014}:
\begin{equation}\label{eq:crit-max-min}
    \begin{split}
        \bm u_* = \argmax_{\bm u\in\mathbb U} & \set{f_{\bm U}(\bm u)^{(\operatorname{dim}(\mathbb U)^{-1})} \times \min_{\bm u'\in\mathscr U}\|\bm u - \bm u'\|} \\
        \text{s.t.} & \; q_{\hat\rho^*,0.1} \leq \hat{\chi}(\bm u) \leq q_{\hat\rho^*,0.9}.
    \end{split}
\end{equation}

This criterion adapts the max-min approach of \citet{lacazeGeneralized2014}, originally designed in the context of traditional reliability analysis -- that is, for estimating failure domains corresponding to a subdomain of scalar functions -- to the method of \citet{bulthuisMultifidelity2020}.
While traditional active learning in reliability analysis focuses on identifying a scalar limit-state, our criterion handles functional outputs over a mesh by targeting an improvement of the auxiliary variable estimate $\hat{\chi}(\bm u)$.
Despite this difference in the target quantity, both approaches share the fundamental objective of sequentially minimizing surrogate uncertainty near the boundaries of the domain of unwanted system behavior.
More precisely, it favors the selection of new inputs close to the level set $\hat\rho^*$ of the function $\bm u \mapsto \hat{\chi}(\bm u)$.
This is performed through the constraint, which ensures that only candidates where the estimated $\hat{\chi}(\bm u)$ are falling within the range bounded by the 10\%- and 90\%-quantile of $\hat{\rho}^*$ are considered.
By selecting such a range rather than a single value, we allow $\hat{\chi}(\bm{u})$ to reach values further from the current estimate of $\hat{\rho}^*$. This loosens the constraint, thereby expanding the feasible domain, which should help prevent overfitting $\hat{\chi}(\bm{u})$ to a potentially inaccurate $\hat{\rho}^*$.

The objective function balances two competing goals:
maximizing the distance to the closest sample in the DoE, and prioritizing regions of high input probability density.
Taking the $\operatorname{dim}(\mathbb U)$-th root of $f_{\bm U}$ avoids the probability density being too small when the dimension of $\mathbb U$ becomes large.

\begin{figure}[!ht]
    \centering
    \includegraphics[width=0.45\textwidth]{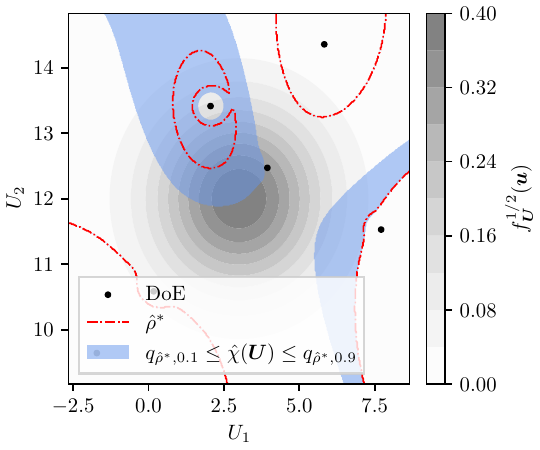}
    \hspace{1cm}
    \includegraphics[width=0.45\textwidth]{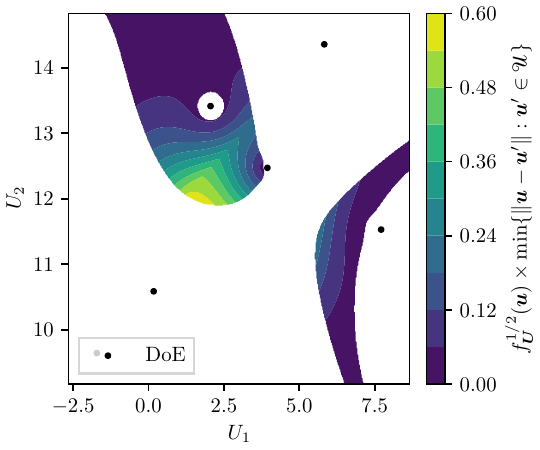}
    \caption{(Left) the different components of \Cref{eq:crit-max-min}: the PDF of $\bm{U}$ is the gray shaded area, the range $q_{\hat{\rho}^*,0.1} \leq \hat{\chi}(\bm{u}) \leq q_{\hat{\rho}^*,0.9}$ is the blue shaded area, the DoE is the black dots, and the level set $\hat{\chi}(\bm{u})=\hat{\rho}^*$ is the red dash-dotted line. (Right) the value of $f_{\bm{U}}(\bm{u})^{\operatorname{dim}(\mathbb{U})^{-1}} \times \min\{\|\bm{u} - \bm{u}'\|:\bm{u}'\in\mathscr{U}\}$ within the range $q_{\hat{\rho}^*,0.1} \leq \hat{\chi}(\bm{u}) \leq q_{\hat{\rho}^*,0.9}$.}
    \label{fig:maxmin-illustrated}
\end{figure}

In \Cref{fig:maxmin-illustrated}, we provide illustrations for an arbitrary function to help understand \Cref{eq:crit-max-min}.
On the left, the PDF of $\bm{U}$ is represented by the shaded gray area, while the blue area depicts the subdomain which satisfies the constraint, that is the range $q_{\hat{\rho}^*,0.1} \leq \hat{\chi}(\bm{u}) \leq q_{\hat{\rho}^*,0.9}$, which is updated at each iteration.
The DoE is shown as black dots and the estimated level set $\hat{\chi}(\bm{u})=\hat{\rho}^*$ is the red dash-dotted line.
On the right, we show the value of the objective function $f_{\bm{U}}(\bm{u})^{\operatorname{dim}(\mathbb{U})^{-1}} \times \min\{\|\bm{u} - \bm{u}'\|:\bm{u}'\in\mathscr{U}\}$ within the constraint feasibility subdomain.
The DoE is again shown as black dots.
If a new sample were added following the active learning strategy, it would be located in the yellow area near $(1.5,12)$.

The quantile range in the constraint of \Cref{eq:crit-max-min} is chosen to balance exploration and exploitation.
First, the range needs to be wide enough to foster exploration within $\mathbb{U}$ by allowing values of $\hat{\chi}(\bm{u})$ distant from the current estimate $\hat{\rho}^*$.
Second, avoiding values close to 0 and 1 favors exploitation by constraining the search space and avoiding sampling in uninformative regions.
In addition, the chosen quantile thresholds must account for the relatively small number $n_\mathrm{rea}$.
In this paper we chose the 0.1- to 0.9-quantile range.
However, we also tested the 0.25- to 0.75-quantile range, which yielded no noticeable difference in the performance of the global algorithm.

At each iteration, the training datasets are updated to $\mathscr U\gets\mathscr U\cup\{\bm u_*\}$ and $\mathscr Y\gets\mathscr Y\cup\{\bm y(\bm u_*)\}$.
The surrogate model is updated accordingly, as are all the quantities described in \Cref{sec:surrogate-estimation}.
The algorithm is stopped when the computational budget $b_\mathrm{max}$ is reached.
This budget is defined by the analyst based on practical engineering constraints, available computational resources, or prior expert knowledge.
A pseudo-algorithm of the active learning process is given in \Cref{alg:excursion}.
To solve the constrained optimization problem, we rely on a sampling-based approach: the acquisition function is evaluated over the entire Monte Carlo population, and the feasible sample yielding the highest value is selected.

\begin{figure}[htb]
\centering
    \begin{algorithm}[H]
        \begin{algorithmic}
            \Require{Initial DoE $(\mathscr{U},\mathscr{Y})$, budget $b_\mathrm{max}$, Monte Carlo samples $\mathscr{U}_\mathrm{MCS}$}
            \Ensure{Confidence region estimate $\hat{\mathcal{C}}_\alpha$}
            \State Train the surrogate from $(\mathscr{U},\mathscr{Y})$
            \State Estimate $\hat{p}$, $\hat{\mathcal{E}}(\bm{u})$, $\hat{\chi}(\bm{u})$ and $\hat{\rho}^*$ using the surrogate, for $\bm{u}\in\mathscr{U}_\mathrm{MCS}$
            \State $k\gets 0$
            \While{$k < b_\mathrm{max}$}
                \State Get $\bm u_*$ from \Cref{eq:crit-max-min}
                \State Run the simulator to get the corresponding output $\bm y(\bm u_*)$
                \State $\mathscr{U}\gets\mathscr{U}\cup \{\bm u_*\}$ and $\mathscr{Y}\gets\mathscr{Y}\cup \{\bm y(\bm u_*)\}$
                \State Train the surrogate from the updated $(\mathscr{U},\mathscr{Y})$
                \State Estimate $\hat{p}$, $\hat{\mathcal{E}}(\bm{u})$, $\hat{\chi}(\bm{u})$ and $\hat{\rho}^*$ using the surrogate, for $\bm{u}\in\mathscr{U}_\mathrm{MCS}$
                \State $k\gets k+1$
            \EndWhile
            \State Estimate $\hat{\mathcal C}_{\alpha}$ with the surrogate as the Vorob'ev quantile $\mathcal Q_{\hat \rho ^*}$ (\Cref{eq:vorobev-quantile})
        \end{algorithmic}
        \caption{Learning excursion-set confidence regions.}
        \label{alg:excursion}
    \end{algorithm}
\end{figure}

\section{Numerical experiments}\label{sec:numerical-experiments}

Under the assumptions of the proposed surrogate framework, the methodology described above is applicable to any deterministic simulator, irrespective of the underlying physics, provided that the mesh does not change between simulator runs. In our work, we focus on two aerospace applications.
In this section, we first introduce the performance metrics used to evaluate the effectiveness of the proposed active learning strategy.
The strategy is then applied to three case studies: a synthetic example, the surface pressure distribution of a hypersonic vehicle, and the return-to-launch-site trajectory of a space launcher first stage.

We compare our method to two alternative strategies.
The first uses a surrogate model trained on a LHS with the same number of training samples as our approach, changing at each iteration to keep the LHS properties.

The second is the method proposed and tested by \citet{bulthuisMultifidelity2020}.
It starts from a few samples drawn from the input distribution at which the simulator is run.
These samples are used to train a KDE model \citep{parzenEstimation1962,wandKernel1994} to approximate the coverage probability function $p$.
From these samples, they compute the corresponding values of the auxiliary variable $\chi(\bm{U})$.
Then, a PCE model \citep{ghanemSpectral1991,xiuWienerAskey2002,soizePhysical2004} maps the input space to $\chi(\bm{U})$.
With the latter, the auxiliary variable is predicted for the entire Monte Carlo population, from which the Vorob'ev threshold $\rho^*$ is estimated.
Iteratively, the authors run the simulator for the sample giving the predicted auxiliary variable closest to the estimated $\rho^*$.
Note that since this new sample is not drawn from the distribution of $\bm U$, it cannot be used to benefit $\hat{p}$, which is thus not updated during the learning process.
For full details of this method, readers are referred to the original paper \citep{bulthuisMultifidelity2020}.

The numerical experiments are implemented in Python.
Our PCA is based on the Scikit-learn PCA \citep{pedregosaScikitlearn2011}, while Kriging and GP sampling are performed with the SMT library \cite{savesSMT2023}.
For the latter, the marginal likelihood is optimized using the Cobyla algorithm \citep{powellDirect1994} with 20 random starting points.
PCE and probabilistic modeling are performed using OpenTURNS \citep{baudinOpen2015}.
For PCE, we use least-angle regression \citep{blatmanAdaptive2011} to identify the sparse estimate of the coefficients.
Other computations are carried out with our own implementations which rely on NumPy \citep{harrisArray2020}.
Details of the numerical settings for the individual case studies are presented when needed.

\subsection{Performance metrics}

To assess the performance of the proposed active learning strategy and compare it with existing methods, we consider three performance metrics.
The first is the effective containment probability: $\hat{\alpha}=\PrU[\mathcal{E}(\bm U)\subset\hat{\mathcal{C}}_{\alpha}]$.
It follows directly from the definition of a confidence region given in \Cref{eq:proba-confidence-region}.
To estimate it, we use the reference excursion sets obtained by running the simulator at the Monte Carlo samples $\mathcal{E}(\bm u)$ for $\bm{u}\in\mathscr{U}_\mathrm{MCS}$, and measure the fraction that are contained in $\hat{\mathcal{C}}_{\alpha}$,

\begin{equation}\label{eq:hat-alpha}
    \hat{\alpha} = \frac{1}{n_\mathrm{MCS}}\sum^{n_\mathrm{MCS}}_{i=1}\mathbf{1}_{\hat{\mathcal{C}}_\alpha}(\mathcal{E}(\bm u_i)),
\end{equation}
with $\mathbf 1_{\hat{\mathcal{C}}_\alpha}$ the indicator function which equals 1 if $\mathcal{E}(\bm u_i)\subset \hat{\mathcal{C}}_\alpha$, and 0 otherwise.
It is important to note that $\hat{\alpha}$ is not available to the end user, since it requires access to the reference excursion sets of the true simulator.

To isolate the effect of the surrogate model and disregard the Monte Carlo approximation error, $\hat{\alpha}$ is not directly compared to the prescribed containment probability $\alpha$.
Instead, we compute the reference confidence region $\mathcal{C}^\mathrm{MCS}_\alpha$ using the true simulator combined with the Monte Carlo samples.
Then, because of the Monte Carlo finite sample, the effective containment probability of $\mathcal{C}^\mathrm{MCS}_\alpha$ is not $\alpha$, but rather $\alpha_\mathrm{MCS}$.
It is obtained by using \Cref{eq:hat-alpha} with $\mathcal{C}^\mathrm{MCS}_\alpha$.

The second performance metric is the volume of the symmetric difference between the reference confidence region and its surrogate-based estimate.
While the first metric evaluates whether the definition of the confidence region is satisfied or not, that is, the containment probability is right or not, this one quantifies the error in terms of volume of the confidence region.
It is computed with
\begin{equation}
    \Vol(\mathcal{C}_\alpha\bigtriangleup \hat{\mathcal{C}}_\alpha)=\Vol\left((\mathcal{C}_\alpha \backslash \hat{\mathcal{C}}_\alpha) \cup (\hat{\mathcal{C}}_\alpha \backslash \mathcal{C}_\alpha)\right),
\end{equation}
where $\bigtriangleup$ denotes the symmetric difference between sets.
Once again, since $\mathcal C_\alpha$ is not available in practice, this metric is inaccessible in practical engineering scenarios.

The third performance metric quantifies two sources of modeling uncertainty associated with the confidence region $\hat{\mathcal C}_{\alpha}$.
The first source is the predictive uncertainty of the GPs.
To evaluate it, we draw $n_\mathrm{rea}$ realizations from the GPs and compute the corresponding realizations of the confidence region $\hat{\mathcal{C}}_{\alpha,\mathrm{GP}_i}$, for $i\in\{1,\dots,n_\mathrm{rea}\}$.
For any mesh node location $\bm x$, the probability of belonging to the confidence region is estimated as
\begin{equation}\label{eq:proba-gp}
    \operatorname{Pr}_{\textrm{GP}}[\bm x\in\hat{\mathcal{C}}_\alpha] = \frac{1}{n_\mathrm{rea}}\sum^{n_\mathrm{rea}}_{i=1}\mathbf{1}_{\hat{\mathcal C}_{\alpha,\mathrm{GP}_i}}(\bm x),
\end{equation}
where $\mathbf{1}_{\hat{\mathcal C}_{\alpha,\mathrm{GP}_i}}(\bm x)=1$ if $\bm x$ belong to the $i$-th realization of the confidence region, and 0 otherwise.

A second source of uncertainty comes from the choice of the initial DoE.
To quantify it, we run the active learning strategy  $n_\mathrm{DoE}$ times with different initial DoE, and estimate the probability that a mesh node location $\bm x$ belongs to the confidence region using
\begin{equation}\label{eq:proba-doe}
    \operatorname{Pr}_{\textrm{DoE}}[\bm x\in\hat{\mathcal{C}}_\alpha] = \frac{1}{n_\mathrm{DoE}}\sum^{n_\mathrm{DoE}}_{i=1}\mathbf{1}_{\hat{\mathcal C}_{\alpha,\mathrm{DoE}_i}}(\bm x),
\end{equation}
with $\mathbf{1}_{\hat{\mathcal C}_{\alpha,\mathrm{DoE}_i}}(\bm x)=1$ if $\bm x$ belongs to the confidence region obtained at the end of the active learning procedure starting from the $i$-th initial design, and 0 otherwise.

Note that in the following sections, some results are shown for a particular initial DoE.
To ensure a fair comparison, this particular DoE is the final one that achieves the median $\hat\alpha$ across the repetitions for all methods.
It is referred to as the \emph{median DoE}.

Finally, regarding computational efficiency, the algorithmic complexity of the proposed methodology at each active learning iteration can be broken down into three main operations. First, the offline training of the PCA-GP surrogate scales as $\mathcal{O}(d_z \times n^3)$ (one GP for each of the $d_{\bm{z}}$ dimensions in the latent space), where $n$ is the training sample size. Second, predicting with the GPs to compute $\hat{\chi}(\bm{u})$ over the Monte Carlo population scales as $\mathcal{O}(d_z \times n_\mathrm{MCS} \times n^2)$ in the sampling approach to optimization. Because the active learning strategy keeps $n$ relatively small, these standard GP related costs remain negligible compared to running the expensive simulator. The most complex operation is estimating the predictive uncertainty on $\hat{\rho}^*$, which requires projecting the latent realizations back into the high-dimensional output space. This inverse mapping scales as $\mathcal{O}(d_y \times d_z \times n_\mathrm{MCS} \times n_\mathrm{rea})$. Given the typically large mesh size $d_y$, Monte Carlo sample size $n_\mathrm{MCS}$ and number of realizations $n_\mathrm{rea}$, this final step constitutes the primary computational bottleneck of the approach.

\subsection{Analytical example -- Sand piles}\label{sec:sand-piles}

This first problem is a synthetic case study.
The spatial vector is $\bm x\in[-2, 2]^2$, and is discretized into an 80-by-80 regular grid, resulting in a total of 6{,}400 mesh nodes.
The uncertain input vector is $\bm U$, follows a multivariate normal distribution:
\begin{equation}
    \bm U\sim \mathcal N\left(
        \begin{bmatrix}
            0 \\
            0
        \end{bmatrix},
        \begin{bmatrix}
            0.25 & 0 \\
            0 & 0.25 \\
        \end{bmatrix}
        \right).
\end{equation}

The simulator output can be viewed as the height of a surface formed by the sum of four sand piles.
Each pile $i\in\{1,2,3,4\}$ is centered at a different location and shaped according to the function
\begin{equation}\label{eq:sand-pile}
    P_i(\bm x,\bm\mu_i,\bm\sigma_i^2)=\frac{1}{2\pi\det(\operatorname{diag}(\bm\sigma_i^2))^{\frac{1}{2}}}\exp\left(-\frac{1}{2}(\bm x - \bm \mu_i)^\top\operatorname{diag}(\bm\sigma_i^2)^{-1}(\bm x - \bm \mu_i)\right),
\end{equation}
where $\operatorname{diag}$ is the vector-to-matrix operator, and with centers and variances specified by $\bm\mu_1=[-3,3]^\top$, $\bm\sigma_1^2=[4,9]^\top$, $\bm\mu_2=[3,3]^\top$, $\bm\sigma_2^2=[9,4]^\top$, $\bm\mu_3=[3,-3]^\top$, $\bm\sigma_3^2=[4,4]^\top$, $\bm\mu_4=[-3,-3]^\top$, $\bm\sigma_4^2=[4,4]^\top$.
The height of each pile nonlinearly depends on the random vector $\bm U$, resulting in a random function.
The simulator output is expressed as a combination of these four piles:
\begin{multline}
    y(\bm u,\bm x) = 1 + 2\sin(3u_1 u_2)P_1(\bm x) + 2u_1^2\exp\left(-\frac{1}{2}u_2^2\right)P_2(\bm x) \\
    +\cos\left(\frac{u_1+u_2}{\pi}\right)P_3(\bm x) +\sin\left(u_1-u_2+\frac{\pi}{3}\right)P_4(\bm x).
\end{multline}

Failure is defined as any location where the simulator output exceeds the threshold $t=1.03$, that is, $\mathbb T=[1.03,\infty)$.
The goal is to construct a confidence region that covers $\alpha=90\%$ of excursion sets.
For the various Monte Carlo estimations conducted, the sample size is $n_\mathrm{MCS}=10{,}000$.
The setting for the surrogate are as follows. PCA is truncated using a RIC of 99.9\%, which results in 4 principal components using an LHS of 100 samples. The Kriging model uses a squared-exponential covariance function optimized by marginal likelihood using Cobyla \citep{powellDirect1994} with 20 restarts and a nugget factor of $10^{-8}$.
The initial DoE consists of $10\times\operatorname{dim}(\mathbb{U})=20$ samples drawn with an LHS strategy within $[-2,2]^2$, and $30\times\operatorname{dim}(\mathbb{U})=60$ additional samples are added using the proposed active learning strategy.

\begin{figure}[!ht]
    \centering
    \includegraphics[height=5.2cm]{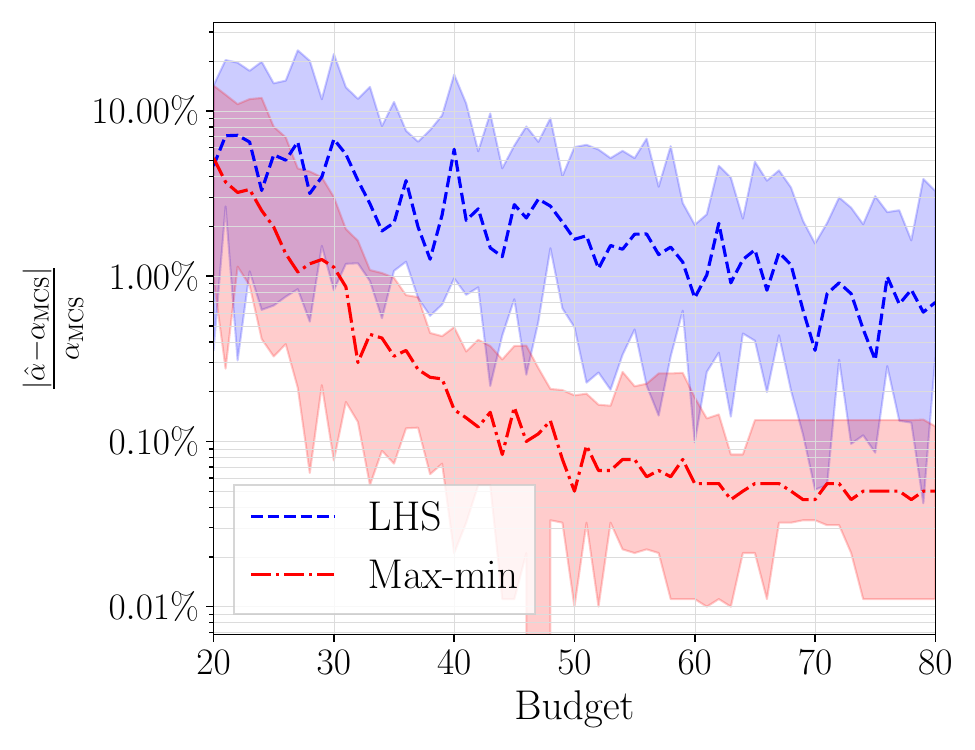}
    \hspace{1cm}
    \includegraphics[height=5.2cm]{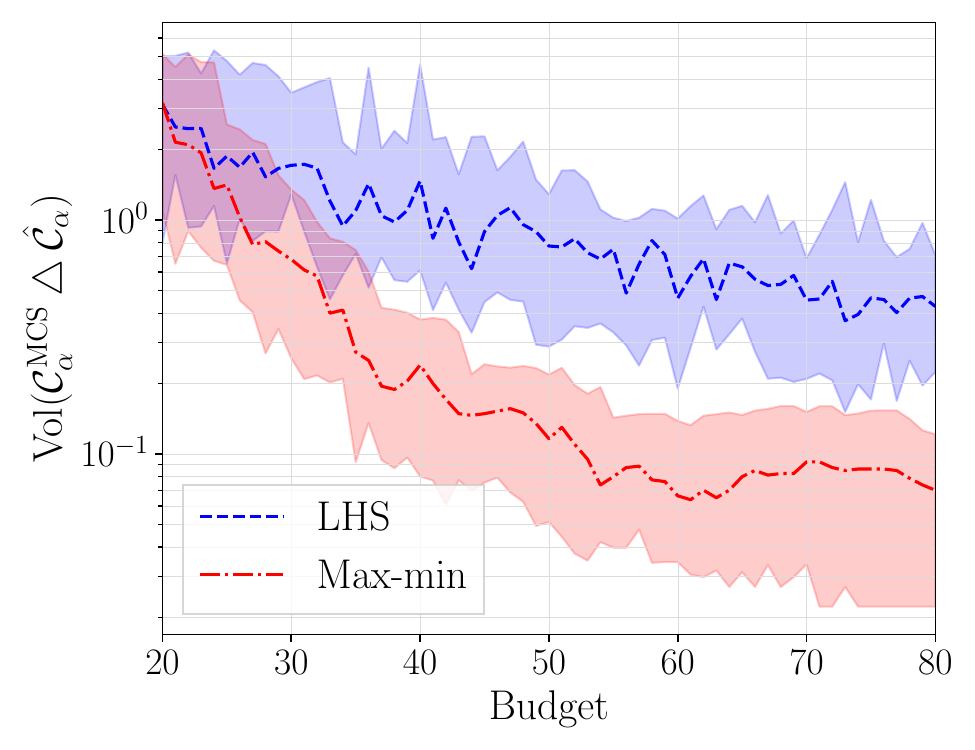}
    \caption{Convergence of $\hat{\alpha}$ (left) and $\Vol(\mathcal{C}_\alpha\bigtriangleup \hat{\mathcal{C}}_\alpha)$ (right) for the synthetic case study. The line is the median across the repetitions and the area between the 0.1- and 0.9-quantile is shaded.}
    \label{fig:norms-a}
\end{figure}

First, we analyze the convergence of the effective containment probability $\hat{\alpha}$.
The results are shown on the left of \Cref{fig:norms-a} in logarithmic scale, where the naive LHS approach is in blue, and our strategy is in red.
The lines represent the median across 20 repetitions, and the shaded area denotes the 0.1- to 0.9-quantile range.
With 20 initial training points, the median relative error is around 5\%, which might seem small initially.
We will illustrate later in the launcher case study of \Cref{sec:launcher} that a small error in $\hat{\alpha}$ can lead to a significantly larger error in the mesh space.

Using our strategy leads to a fast convergence rate for the first 50 iterations or so.
The median error on $\hat{\alpha}$ reaches approximately 0.05\% around iteration 60, and then stabilizes.
Overall, it decreases by two orders of magnitude from start to end.
In comparison, the naive space-filling approach achieves a relative median error around 0.5\% by the end of the learning process, one order of magnitude larger than that provided by our method.

Next, the assessment of the volume of the symmetric difference between the reference confidence region and its estimate shown on the right of \Cref{fig:norms-a} in logarithmic scale, provides insight into the error incurred in the mesh space.
The convergence rate of this metric is comparable to that of the effective containment probability.
In both cases, the volume of the symmetric difference between the reference confidence region and its estimate is around 3 (no unit, representing around 19\% of the total mesh) at the beginning.
By the end of the algorithm, the LHS approach reaches a volume around 0.4\%.
In contrast, with the our strategy, the rate of decrease is significantly faster.
The volume reaches 0.07\% by iteration 60, and stabilizes afterwards.

\begin{figure}[!ht]
    \centering
    \includegraphics[height=5.2cm]{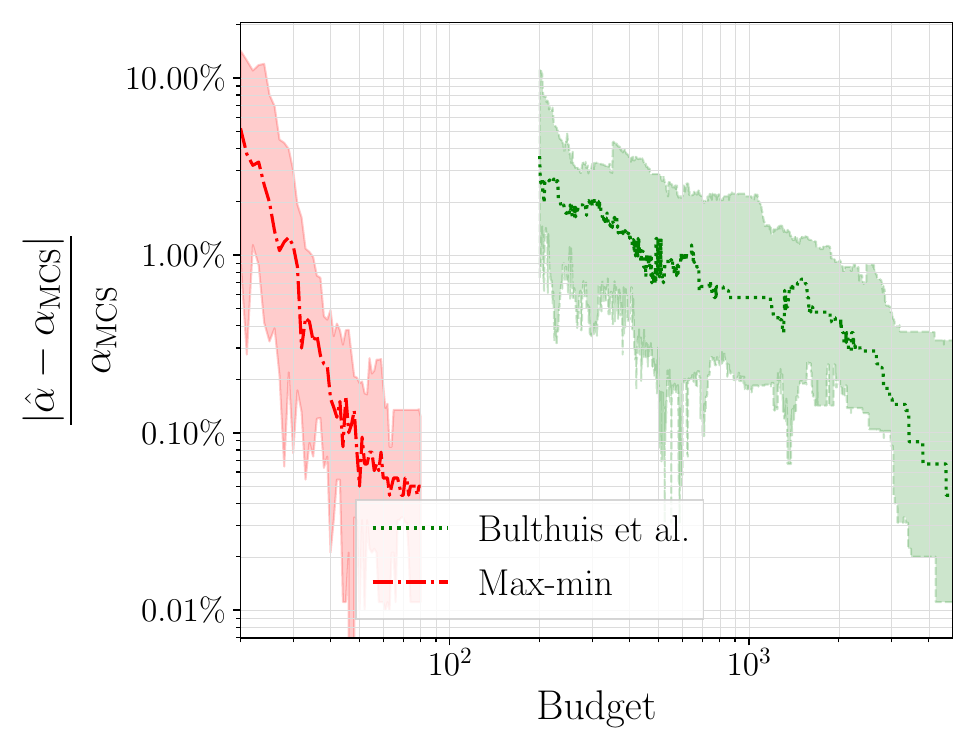}
    \hspace{1cm}
    \includegraphics[height=5.2cm]{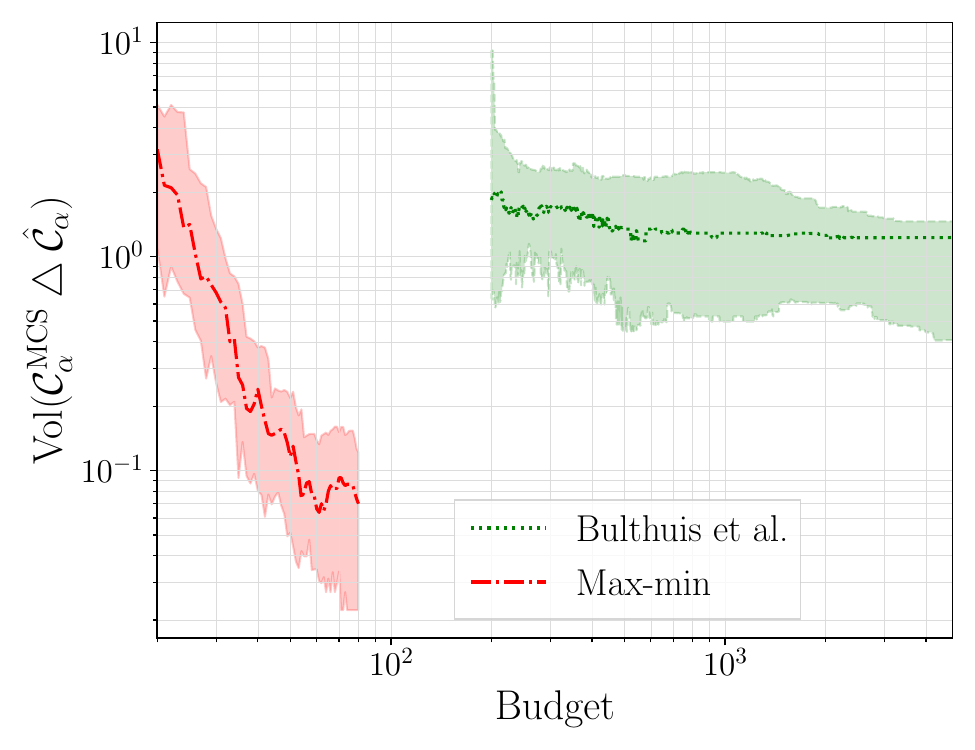}
    \caption{Convergence of $\hat{\alpha}$ (left) and $\Vol(\mathcal{C}_\alpha\bigtriangleup \hat{\mathcal{C}}_\alpha)$ (right) for the synthetic case study, for our method in red, and the one proposed by \citet{bulthuisMultifidelity2020} in green. The line is the median across the repetitions and the area between the 0.1- and 0.9-quantile is shaded.}
    \label{fig:norms-bulthuis}
\end{figure}

A comparison with the method proposed by \citet{bulthuisMultifidelity2020} is presented in \Cref{fig:norms-bulthuis}.
On the left, we show the convergence of $\hat{\alpha}$.
Since their approach relies on a KDE based on the initial training samples to approximate $p$, it requires a relatively large number of initial samples to achieve an error level comparable to that of our strategy.
In this particular case study, their method ultimately demands a significantly larger number of learning samples (by several orders of magnitude) to reach an accuracy comparable to our method.
Precisely, it starts with a median relative containment probability error around 3\% for 200 samples, and reaches 0.04\% with 5{,}000 points -- half the number of Monte Carlo samples, whereas our method starts with 20 samples and is considered converged with 80 samples.

The convergence of $\Vol(\mathcal{C}_\alpha^\mathrm{MCS}\bigtriangleup\hat{\mathcal{C}}_\alpha)$ for the method of \citet{bulthuisMultifidelity2020} is shown in green on the right panel of \Cref{fig:norms-bulthuis}.
Contrary to $\hat{\alpha}$, it does not converge: the geometry of the confidence region estimate remains relatively inaccurate despite the many training samples.
The volume of this symmetric difference decreases to just above 6\% of the mesh volume for 5{,}000 training samples.
This can be partly explained by the way the coverage probability function $p$ is approximated.
During the active learning proposed by \citet{bulthuisMultifidelity2020}, the estimated values $\hat{\chi}(\bm{U})$ are replaced with their true value, leading to an increase in the accuracy of $\hat{\alpha}$ up to a small estimation error.
Conversely, as already noted, the KDE model $\hat{p}$ does not get updated along this process.
As a result, since the definition of the confidence region estimate directly relies upon $\hat{p}$, the accuracy is limited by the initial KDE approximation.
In other words, a region respecting the containment probability $\Pr[\mathcal{E}(\bm{U})\subset\hat{\mathcal{C}}_\alpha]\geq\alpha$ is found, but it is not the one we are seeking: hence the remaining symmetric difference volume.
To help better understand this performance difference, we provide a more detailed analysis of $\chi$ across $\mathbb{U}$ in \Cref{app:chi}.

\begin{figure}[!ht]
    \centering
    \includegraphics[width=0.35\textwidth]{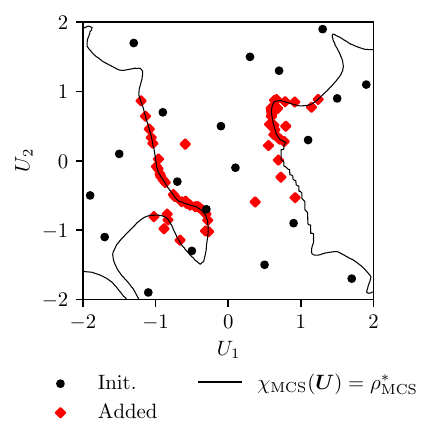}
    \includegraphics[width=0.35\textwidth]{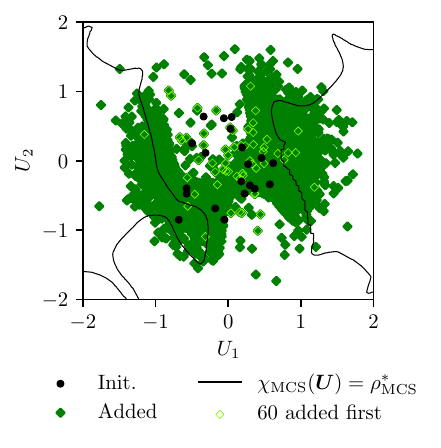}
    \caption{Initial and added samples locations in $\mathbb{U}$ for the sand-piles case with the proposed strategy on the left, and the method from \citet{bulthuisMultifidelity2020} on the right.}
    \label{fig:pairplot-excursion-normsfast}
\end{figure}

The location of the training samples are shown in \Cref{fig:pairplot-excursion-normsfast} for the median repetition.
The max-min criterion is shown on the left, and the approach by \citet{bulthuisMultifidelity2020} on the right.
In the left figure, we see that our approach places training samples very close to the level set $\chi(\bm{u})=\rho^*$, which explains the accuracy in the final estimate.
Conversely, in the right figure, the first 80 iterations of the method of \citet{bulthuisMultifidelity2020} do not seem to follow any obvious pattern.
The level set is eventually identified as later samples fall around it.

\begin{figure}[!ht]
    \centering
    \begin{tabular}{ll}
        \;\includegraphics[width=0.31\linewidth]{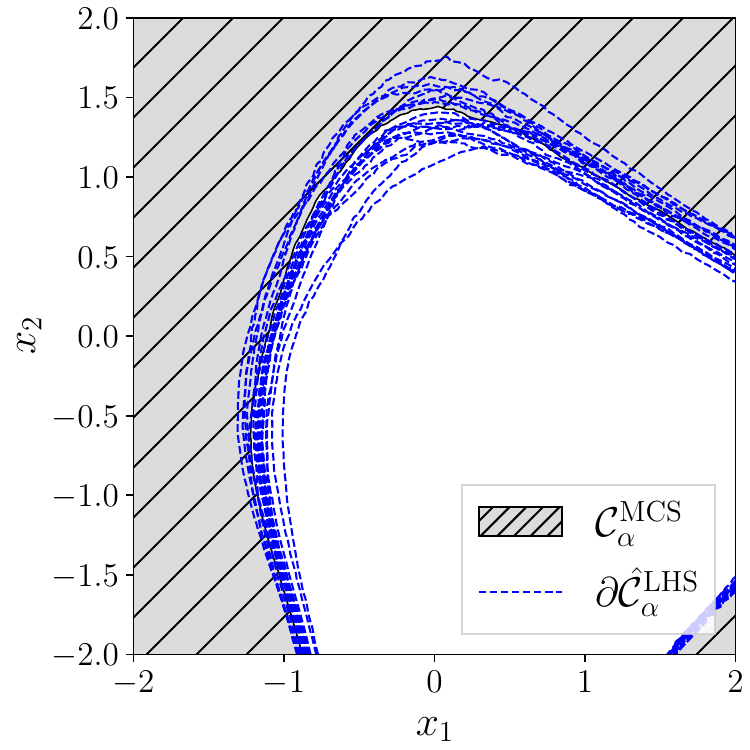} & \;\includegraphics[width=0.31\linewidth]{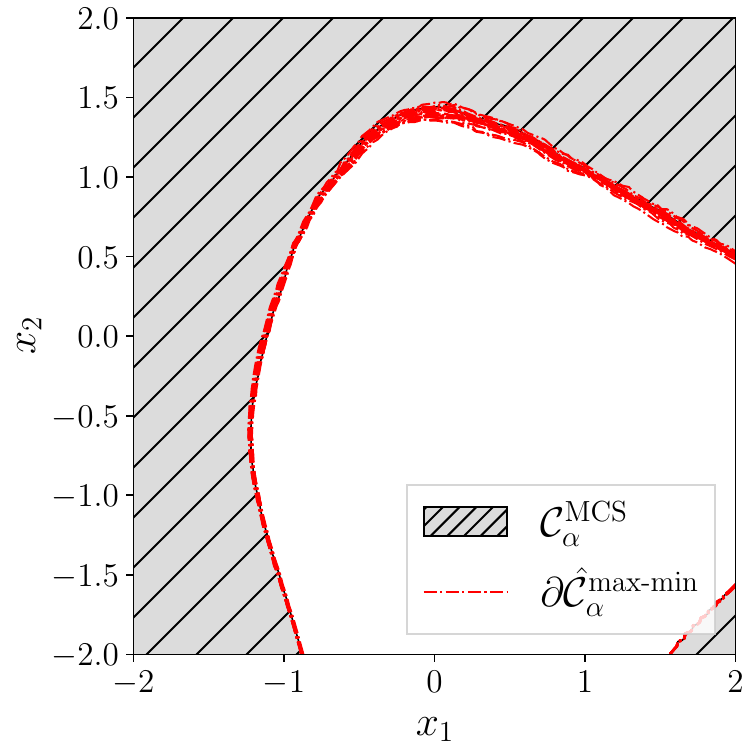} \\
        \includegraphics[width=0.4\linewidth]{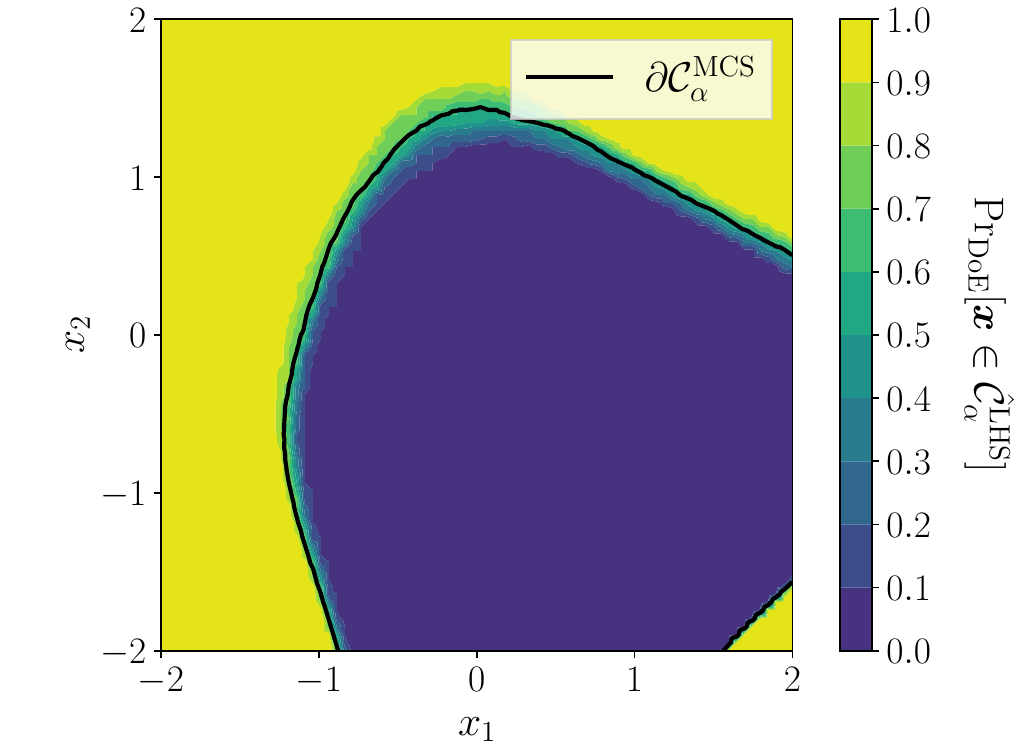} & \includegraphics[width=0.4\linewidth]{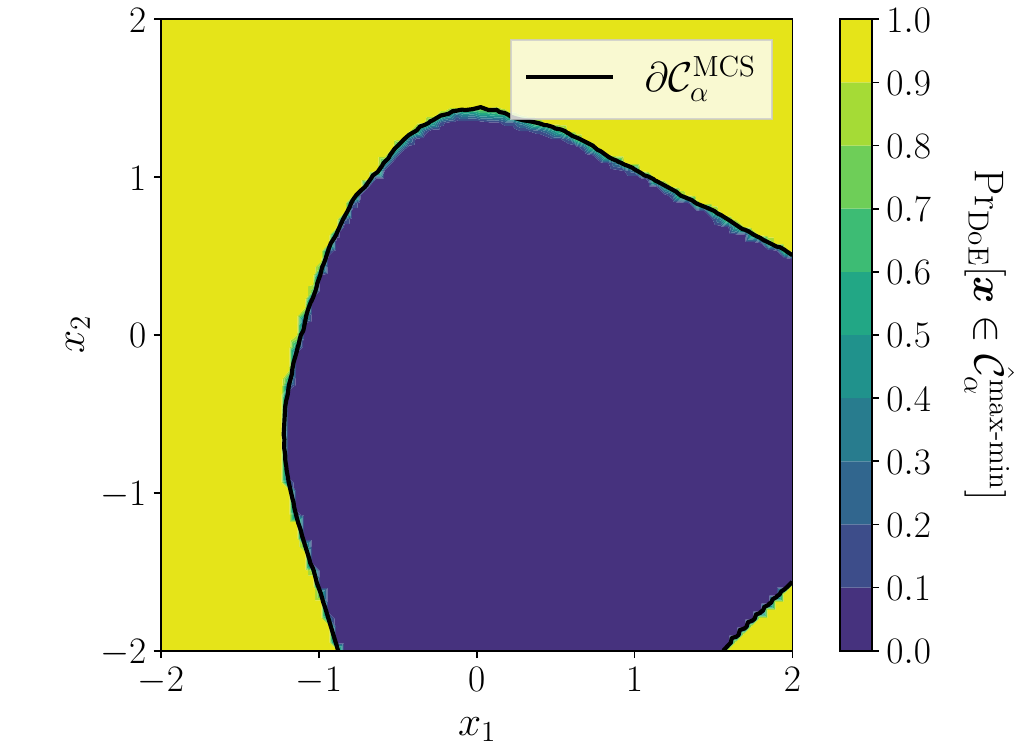}
    \end{tabular}
    \caption{Variability of the estimated confidence region to the choice of the initial LHS. Top: boundaries of the confidence region obtained with the naive LHS (left) and our active learning strategy (right) across 20 repetitions. Bottom: probability that each mesh node belongs to the estimated confidence region, computed from the 20 repetitions using the naive LHS (left) and our approach (right). The boundary of the reference confidence region is in black.}
    \label{fig:norms-C-doe}
\end{figure}

Another important metric is the variability of the estimated confidence region with respect to the choice of initial DoE.
The top row of \Cref{fig:norms-C-doe} shows the confidence regions obtained across 20 independent repetitions, using the naive LHS approach (left, in blue) and our proposed method (right, in red).
The naive approach exhibits significantly higher variability, while our strategy produces consistent results, although a small bias can be seen in the uppermost values of $x_2$.
A more quantitative illustration is provided in the bottom row of \Cref{fig:norms-C-doe}, which shows the probability $\operatorname{Pr}_{\textrm{DoE}}[\bm x\in\hat{\mathcal{C}}_{\alpha}]$ defined in \Cref{eq:proba-doe}.
The result confirms the previous observation: with the proposed strategy, most mesh locations have a probability close to either 0 or 1, indicating consistent confidence regions boundaries across different initial designs.
Meanwhile, with the LHS strategy, the transition between probabilities of 0 or 1 occupies a significant portion of the mesh.
The low variability indicates that, in this case study, our method is not affected by the choice of the initial LHS.

\begin{figure}[!ht]
    \centering
    \begin{tabular}{ll}
        \;\includegraphics[width=0.31\linewidth]{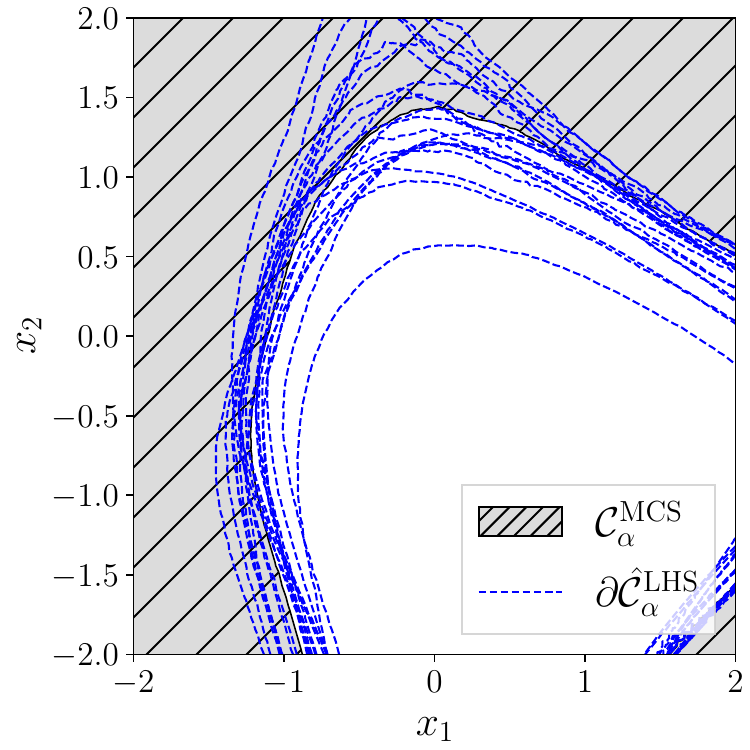} & \;\includegraphics[width=0.31\linewidth]{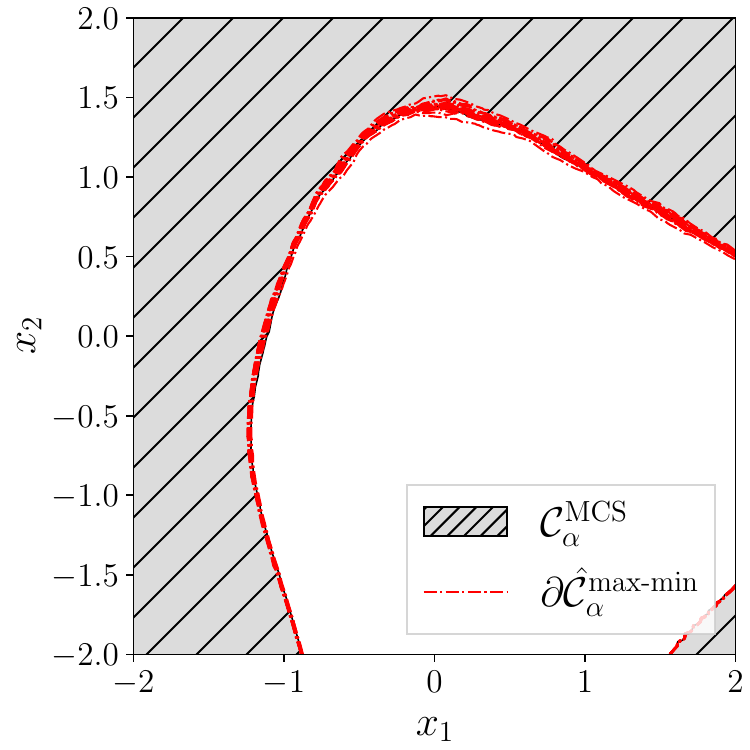} \\
        \includegraphics[width=0.4\linewidth]{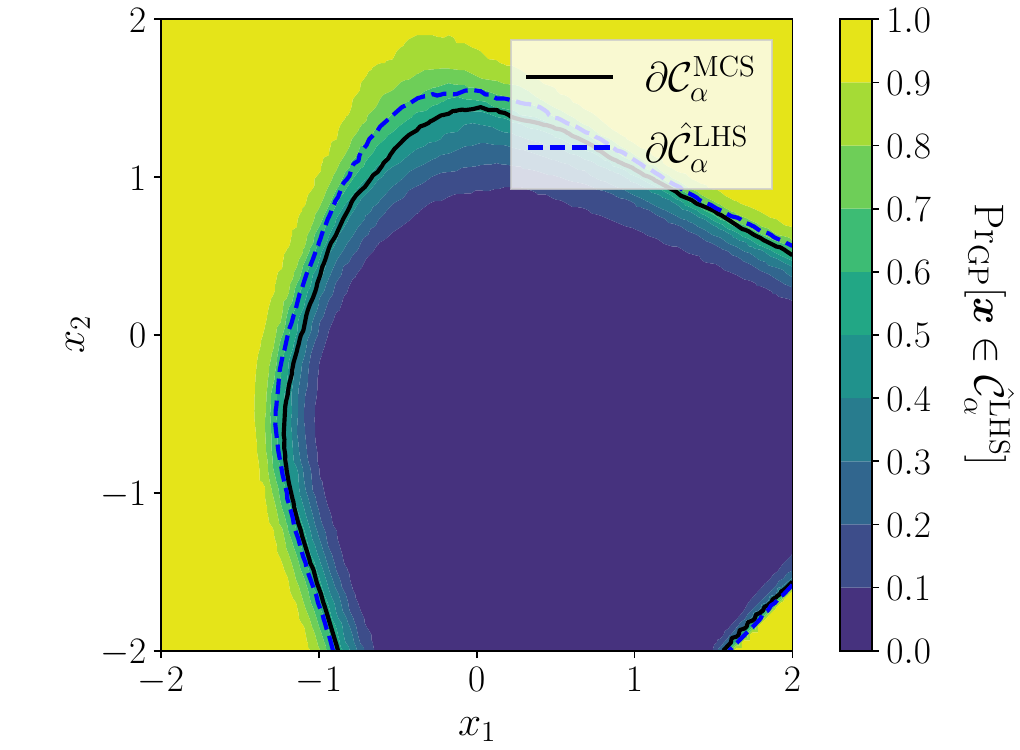} & \includegraphics[width=0.4\linewidth]{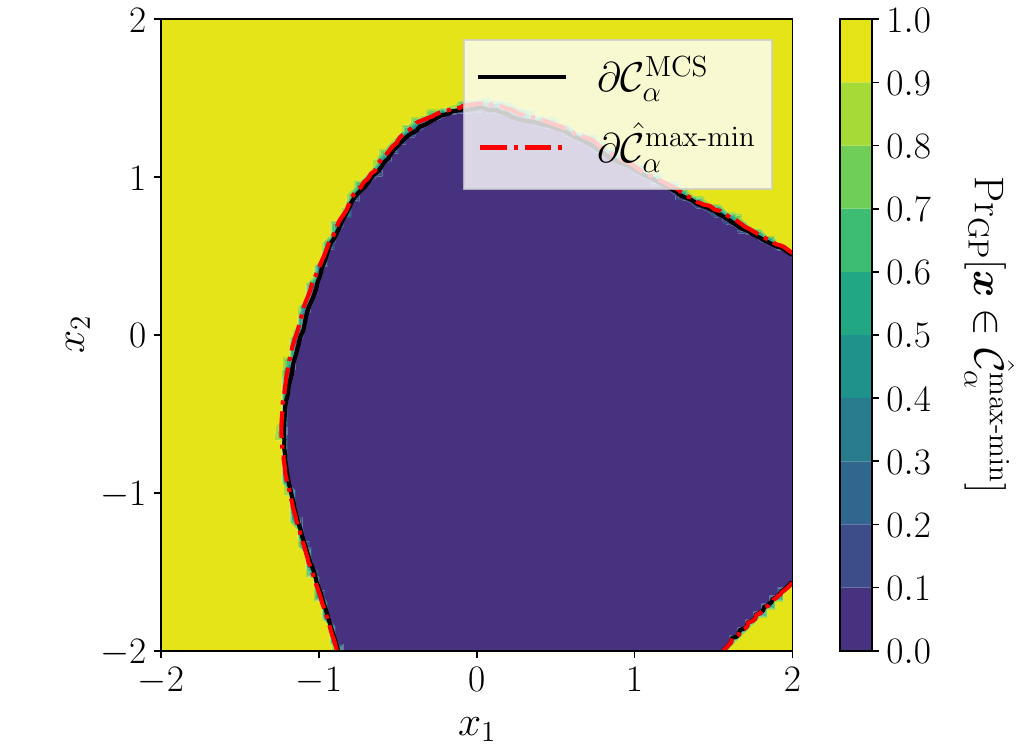}
    \end{tabular}
    \caption{Variability of the estimated confidence region to the intrinsic variability of the GPs for the median DoE. Top: boundaries of the confidence region obtained with the naive LHS (left) and our active learning strategy (right) across 20 realizations of the GPs. Bottom: probability that each mesh node belongs to the estimated confidence region, computed from 200 realizations using the naive LHS (left) and our approach (right). The boundary of the reference confidence region is in black.}
    \label{fig:norms-C-gp}
\end{figure}

The final metric of interest concerns the impact of the intrinsic variability of the GPs on the estimated confidence region.
The top row of \Cref{fig:norms-C-gp} displays 20 realizations of the confidence region, each obtained by drawing a different realization of the latent GPs, all for the median DoE.
As for the previous metric, our learning strategy results in substantially lower variability compared to the naive LHS approach.
This is particularly important because it is the only accessible metric in practical applications where the simulator is costly.
The bottom row of \Cref{fig:norms-C-gp} shows the probability of each mesh node of being included in the confidence region, accounting the uncertainty of the GPs.
It is estimated using 200 realizations of the latent GPs.
The conclusion remains the same: our method yields a crisp estimated confidence region in the sand-piles case, while for the space-filling approach this transition zone spans a significant portion of the mesh.

\subsection{Hypersonic vehicle}\label{sec:hypersonic}

\begin{figure}[!ht]
    \centering
    \includegraphics[width=\textwidth]{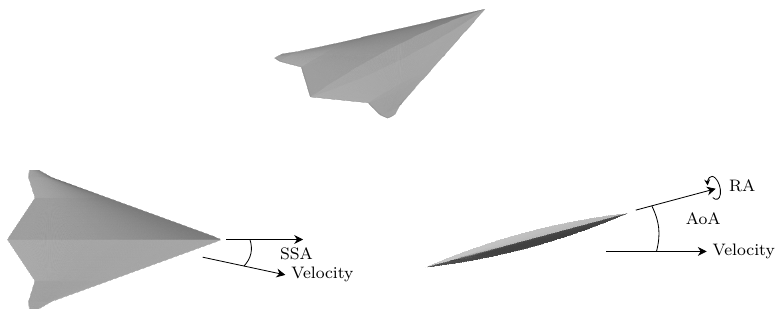}
    \caption{Three-dimensional, top, and right views of the hypersonic vehicle. AoA: angle of attack; RA: roll angle; SSA: side-slip angle. Some angles are exaggerated to improve readability.}
    \label{fig:supersonic}
\end{figure}

The first physical case study focuses on the design of a hypersonic vehicle.
Due to their extremely high velocities, hypersonic vehicles are subject to intense aerodynamic loads, resulting in significant stresses on the vehicle structure.
In particular, the walls of the vehicle can experience extremely high pressures during flight.
Consequently, the design of the vehicle is critical to ensure structural integrity and mission success, while preventing excessive mass.
Additionally, this information could support further studies, such as actuator sizing for vehicle control, and shock-wave analysis for engine air intakes.
A three-dimensional view of the vehicle is shown in the top of \Cref{fig:supersonic}.

To evaluate the aerodynamic environment, we use an in-house simulation tool named SHAMAN, which implements local surface inclination methods \citep{rolimDevelopment2020,leeEfficient2021}.
It computes the pressure coefficient field (no unit) along the vehicle surface, which is used to identify regions likely to experience extreme loads.
The computational domain is discretized using a surface mesh composed of 159{,}184 cells.
However, for this study, we restrict our attention to the pressure field on the 80{,}384 cells of the vehicle \emph{lower surface}.
This surface field is projected onto a two-dimensional plane to facilitate the analysis and visualization.
This case study demonstrates the scalability of our methodology to larger meshes.

The vehicle is flying with a high angle of attack, approximately 40$^\circ$, which can be assimilated to an atmospheric reentry.
We consider hypersonic velocities around Mach~8.
Uncertainty in the simulator inputs arises from the variability in flight parameters, specifically: the angle of attack $\mathrm{AoA}\sim \mathcal N(40,3^2)$ (in degrees), the roll angle $\mathrm{RA}\sim \mathcal N(0,2^2)$ (in degrees), the side-slip angle $\mathrm{SSA}\sim \mathcal N(0,2^2)$ (in degrees), and the Mach number $\mathrm{M}\sim \mathcal N(8,0.1^2)$, see \Cref{fig:supersonic} for an illustration of the different angles.
The input vector is thus $\bm U=[\mathrm{AoA}, \mathrm{RA}, \mathrm{SSA},\mathrm{M}]^\top$.

The proposed active learning strategy is run for a threshold $t=0.0145$ with thus $\mathbb{T}=[0.0145,\infty)$, and a containment probability $\alpha=95\%$.
The Monte Carlo is composed of $n_\mathrm{MCS}=30{,}000$ samples.
The PCA is truncated with a RIC of 99.9\%, (for reference, yielding 8 principal components using an LHS of 100 sample).
The Kriging model is trained by maximizing the marginal likelihood using Cobyla \citep{powellDirect1994} with 20 restarts.
The covariance model is Mat\'ern-5/2 with a nugget factor of $10^{-10}$.
The initial DoE is composed of $2\times\dim(\mathbb U)=8$ samples and $6\times\dim(\mathbb{U})=24$ are added according to the active learning criterion, across 10 independent repetitions.

\begin{figure}[!ht]
    \centering
    \includegraphics[width=0.55\textwidth]{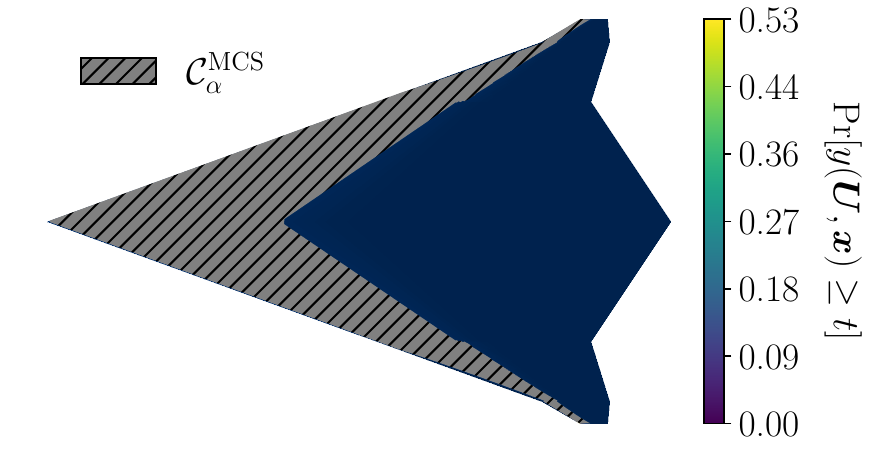}
    \caption{Monte Carlo estimate of the confidence region of the hypersonic vehicle case estimated using the true simulator.}
    \label{fig:supersonic-confidence-region}
\end{figure}

The reference confidence region computed from the Monte Carlo samples is displayed in \Cref{fig:supersonic-confidence-region}.
The estimates are not shown in the following figures because the difference with the reference region is imperceptible.

\begin{figure}[!ht]
    \centering
    \includegraphics[height=5.2cm]{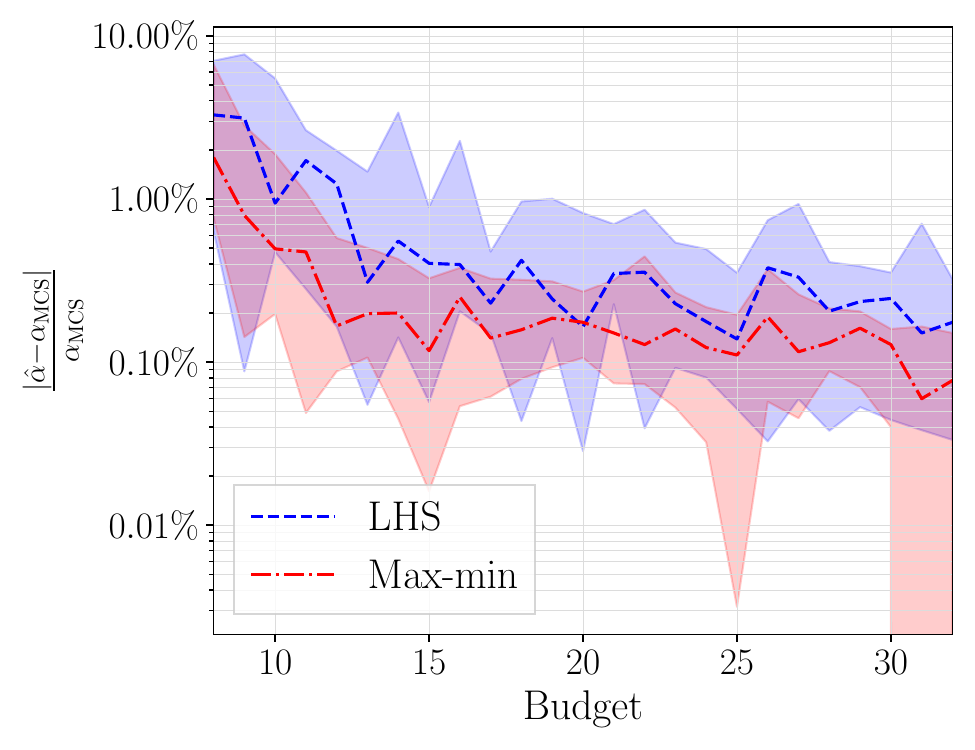}
    \hspace{1cm}
    \includegraphics[height=5.2cm]{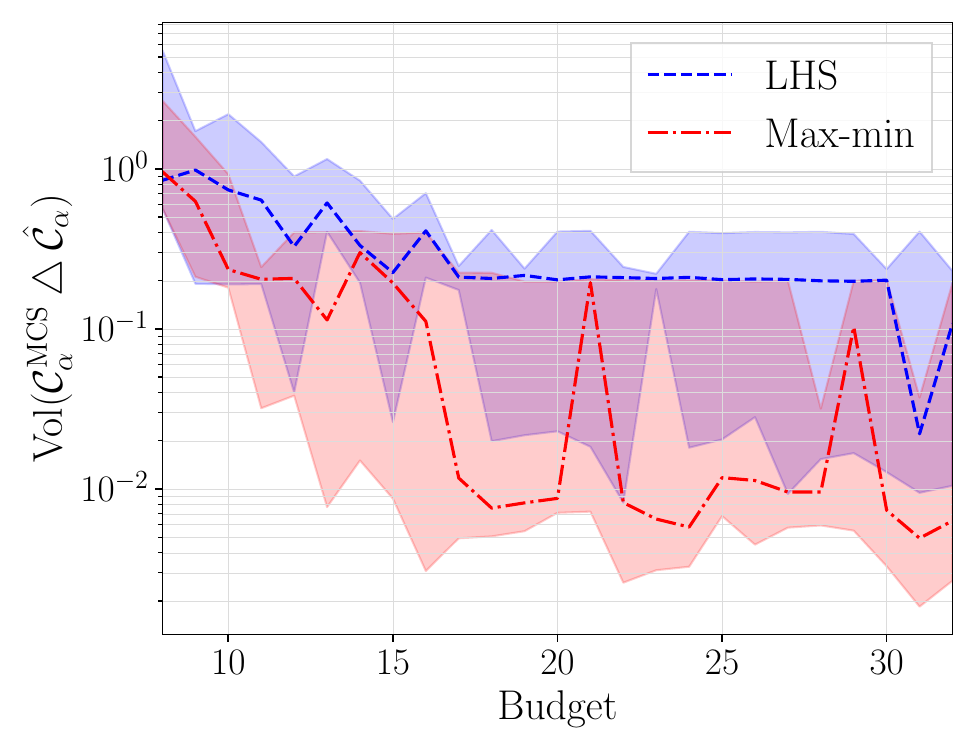}
    \caption{Convergence of $\hat{\alpha}$ (left) and of the volume of the symmetric difference (right) for the hypersonic vehicle case study. The line is the median across the repetitions and the area between the 0.1- and 0.9-quantile is shaded.}
    \label{fig:superso-a-symdiff}
\end{figure}

The relative error in the effective containment probability $\hat{\alpha}$ throughout the learning process is shown on the left of \Cref{fig:superso-a-symdiff} in logarithmic scale.
With only 8 initial samples to train the surrogate, the starting relative error is approximately 3\% for both the naive LHS (blue) and the max-min (red) strategies.
Both methods show similar converge rates.
After 24 additional training samples have been added, the median relative error in $\hat\alpha$ drops to around 0.2\% for the LHS strategy, and approximately 0.07\% for our method, less than half of that of the naive approach.
Both strategies show relatively comparable levels of uncertainty in the estimated $\hat{\alpha}$ across the learning process.
These results suggest that, in scenarios involving relatively smooth physical behavior, the max-min approach still performs better than the simpler LHS design.

The volume of the symmetric difference between the reference confidence region and its estimate, an indicator of error measured in the physical mesh space, is shown on the right of \Cref{fig:superso-a-symdiff} in logarithmic scale.
Initially, both the LHS and max-min strategies yield a volume of the symmetric difference around 0.9 m$^2$.
Considering that the total volume of the mesh is about 27.06 m$^2$, this represents just above 3\% of the domain, suggesting that the initial confidence region estimate is already reasonably accurate due to the smooth physics.
A notable performance gap between the LHS and max-min strategies emerges between around 17 total samples.
The max-min converges more quickly, reaching a volume of the symmetric difference of about 0.007 m$^2$ by the 18th sample and then stabilizing except for two spikes at a budget of 21 and 29, likely caused by the limited number of repetitions.
In contrast, the LHS continues to improve more slowly, stabilizing at around 17 samples, then decreases again near the end, with a final volume of approximately 0.1 m$^2$, about 14 times more than our approach.
It is important to note that the performance of both strategies is ultimately constrained by the mesh discretization itself.

\begin{figure}[!ht]
    \centering
    \includegraphics[height=5.2cm]{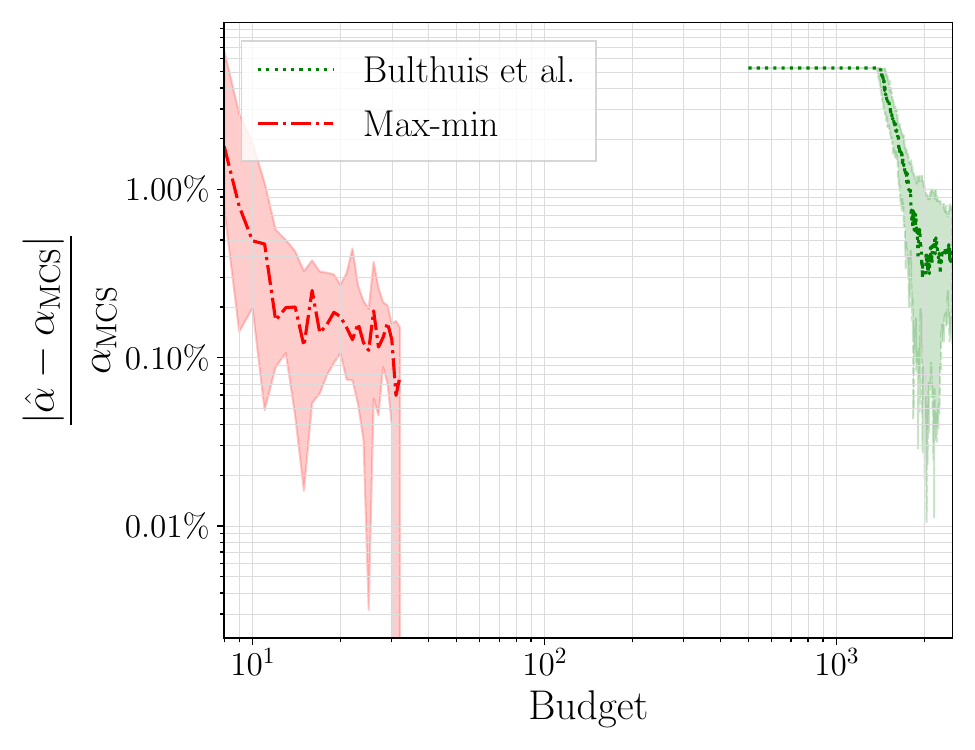}
    \hspace{1cm}
    \includegraphics[height=5.2cm]{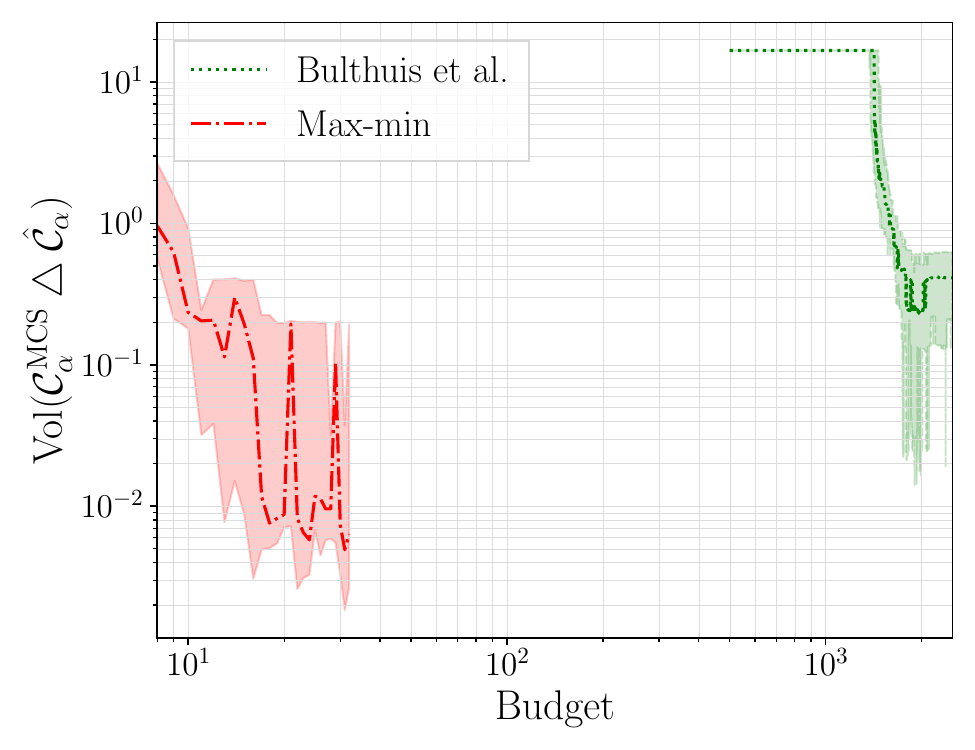}
    \caption{Convergence of the effective containment probability $\hat{\alpha}$ (left) and volume of the symmetric difference (right) for the hypersonic vehicle case study, for our method in red, and the one proposed in \citet{bulthuisMultifidelity2020} in green. The line is the median across the repetitions and the area between the 0.1- and 0.9-quantile is shaded.}
    \label{fig:superso-bulthuis}
\end{figure}

In \Cref{fig:superso-bulthuis}, we compare our method (in red) to that of \citet[][in green]{bulthuisMultifidelity2020}.
On the left, their strategy begins with a relative error in $\hat{\alpha}$ of about 5\% using 500 training points -- more than twice that of our approach using only 8 training points.
The error then stagnates for roughly 600 additional samples before decreasing, eventually reaching approximately 0.4\% with a total of 2{,}500 training samples.
On the right, the symmetric difference between the reference region and its estimate initially exceeds 10~m$^2$ (around 37\% of the total mesh volume), whereas our approach achieves 0.9~m$^2$ already with a budget of 8 points.
Their estimate also begins to improve only after roughly 600 additional points.
By the end of the algorithm, it reaches a symmetric-difference volume of around 0.3~m$^2$ after 2{,}500 simulator evaluations, corresponding to more than 1\% of the mesh domain.

\begin{figure}[!ht]
    \centering
    \includegraphics[width=0.45\textwidth]{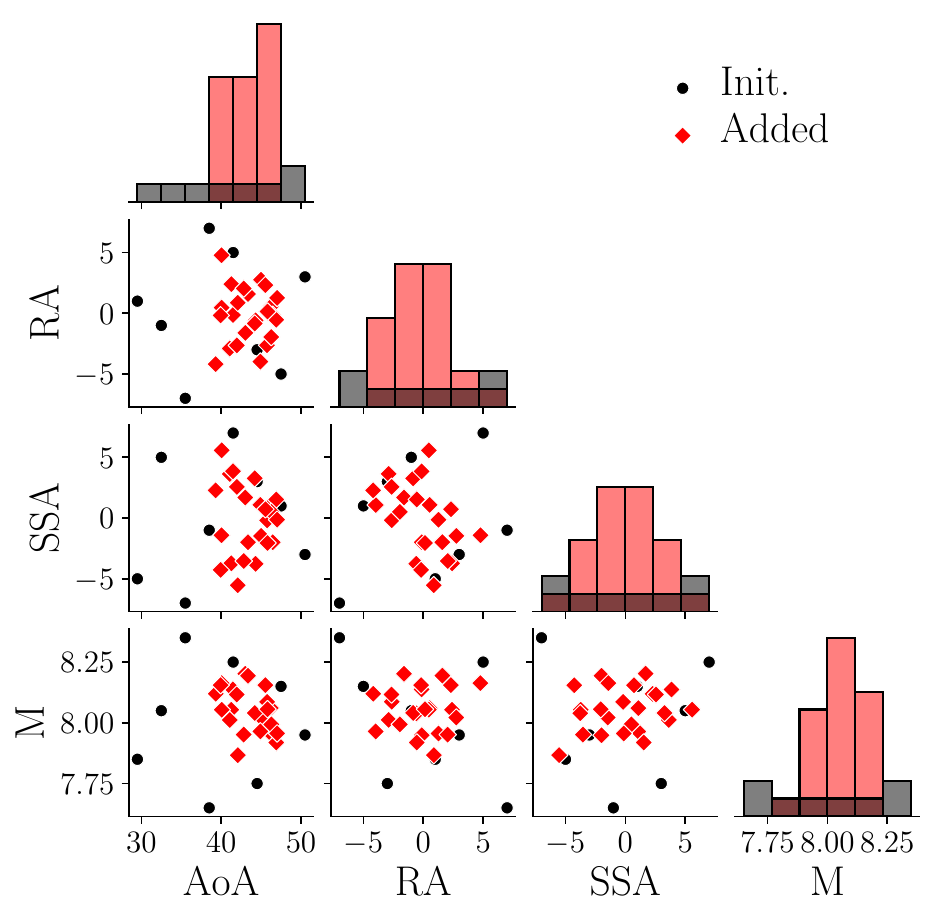}
    \hspace{0.5cm}
    \includegraphics[width=0.45\textwidth]{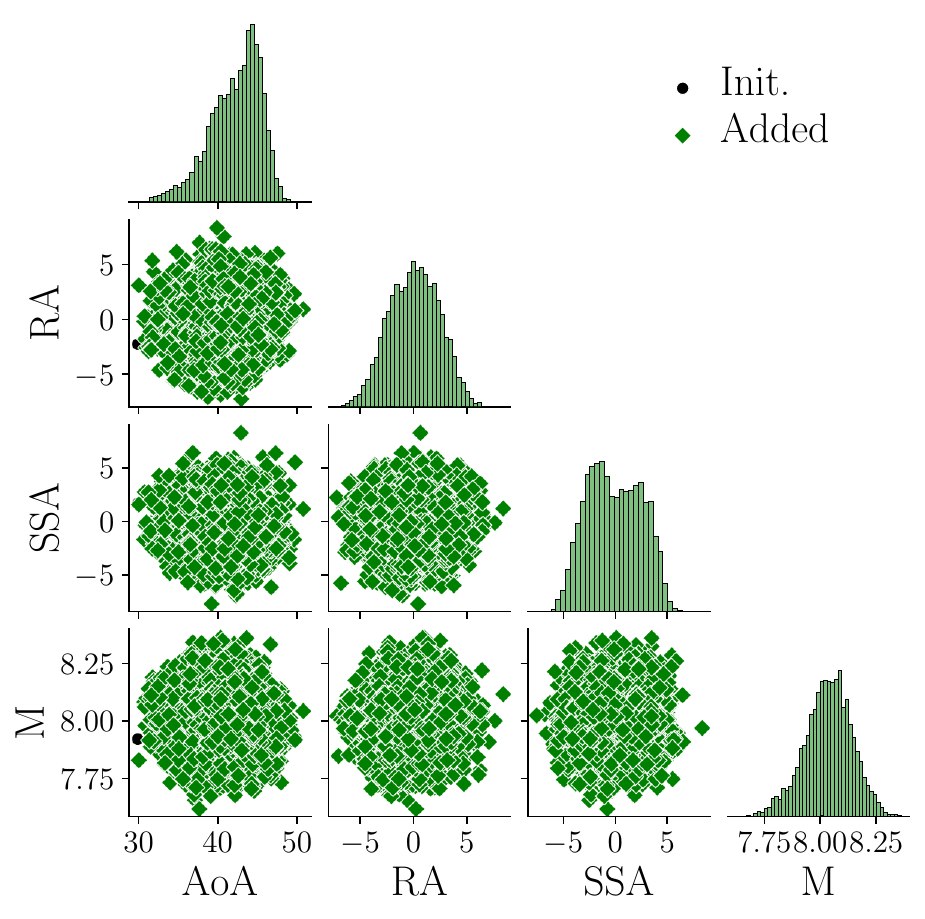}
    \caption{Initial and added samples locations in $\mathbb{U}$ for the hypersonic case, for our method on the left and the strategy of \citet{bulthuisMultifidelity2020} on the right.}
    \label{fig:pairplot-excursion-supersonic}
\end{figure}

The location of the training points in the input domain are shown in \Cref{fig:pairplot-excursion-supersonic} for the median repetition using the max-min criterion on the left, and for the method of \citet{bulthuisMultifidelity2020} on the right.
Due to the four-dimensional nature of the input, it is difficult to notice any clear level set $\chi(\bm{U})=\rho^*$, contrary to the previous case study.
Returning to the physics, points with larger angle of attack and Mach numbers tend to be added to the experimental design in both cases.
The selected roll and side-slip angles are more distributed.
Because the domain is four-dimensional, no clear level set is visible.

\begin{figure}[!ht]
    \centering
    \includegraphics[width=0.45\linewidth]{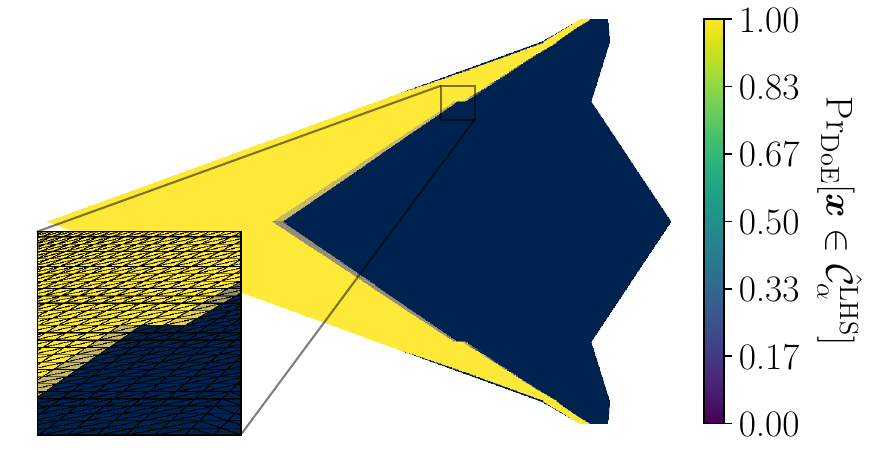}
    \includegraphics[width=0.45\linewidth]{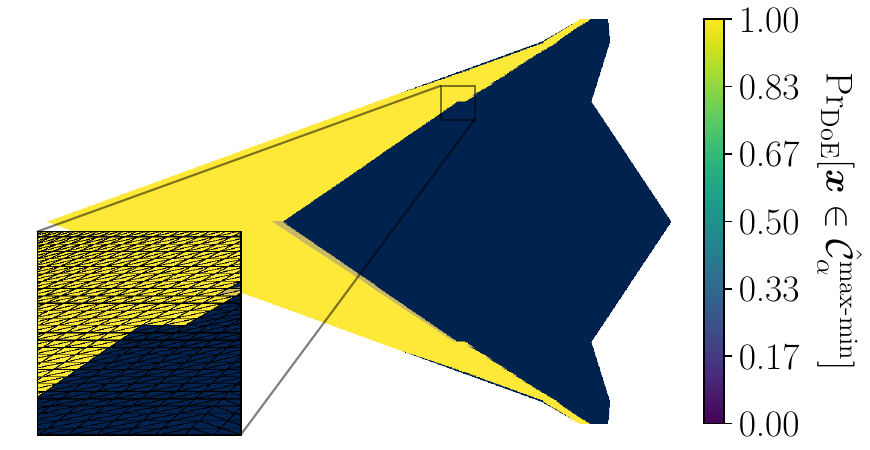}
    \caption{Probability that each mesh node belongs to the estimated confidence region, computed from the 20 repetitions using the naive LHS (left) and our approach (right) in the hypersonic case study.}
    \label{fig:superso-C-doe}
\end{figure}

The variability of the confidence region estimate due to the choice of the initial DoE is illustrated in \Cref{fig:superso-C-doe}.
The left panel shows the probability that a mesh node belongs to the confidence region estimate when the initial DoE is considered random for the naive LHS strategy, while the right panel shows the same for our approach.
In both cases, the influence of the initial DoE appears minimal, likely due to the smoothness of the underlying physics in this case : the transition zone where the probability is neither almost equal to 0 nor 1 is limited to approximately one mesh cell in width.
It is also partly explained by the small number of realizations used to estimate the probabilities, which limits precision.
Overall, these results suggest that, in similar settings, the specific choice of the initial DoE is not a critical factor for either strategy.

\begin{figure}[!ht]
    \centering
    \includegraphics[width=0.45\linewidth]{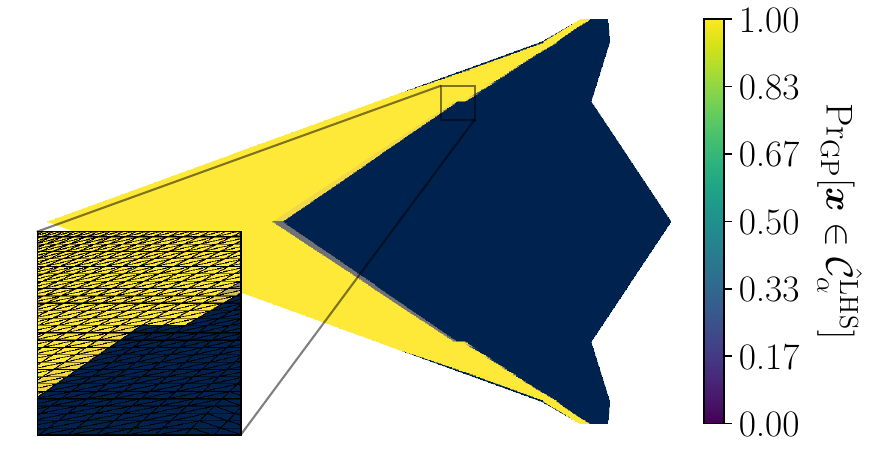}
    \includegraphics[width=0.45\linewidth]{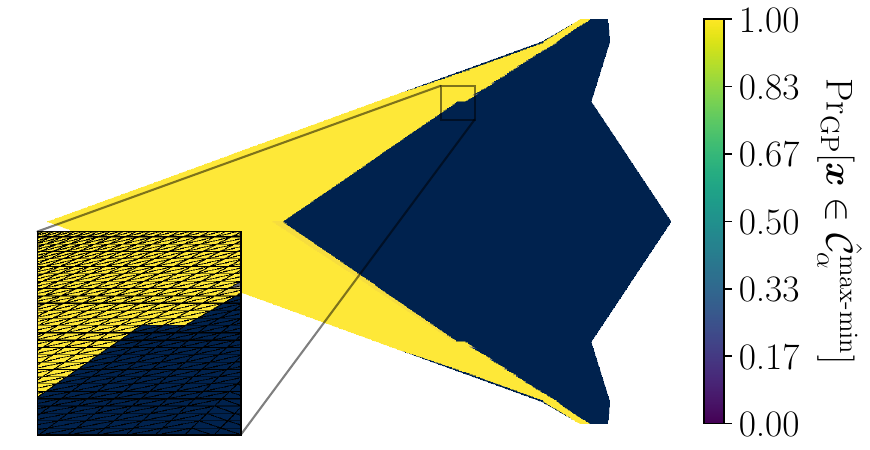}
    \caption{Probability that each mesh node belongs to the estimated confidence region, computed from 200 realizations using the naive LHS (left) and our approach (right) in the hypersonic case study.}
    \label{fig:superso-C-gp}
\end{figure}

Finally, the variability in the confidence region estimate due to the intrinsic uncertainty of the GPs is illustrated in \Cref{fig:superso-C-gp}.
This figure shows the estimated probability that each mesh node belongs to the confidence region, taking into account the posterior distribution of the GP, using the median DoE as training data.
The ability to draw a large number of GP realizations enables a finer and more robust estimation of this probability compared to the DoE-linked one.
Both the naive approach (left) and max-min strategy (right) show similar levels of predictive uncertainty: there is virtually no transition zone (only one mesh in width at worst), and the confidence region boundary is sharply defined, with probabilities close to 0 or 1 across the whole domain.

\subsection{Return-to-launch-site of the first stage of a space launcher}\label{sec:launcher}

The last case study concerns the trajectory of a two-stage-to-orbit launch vehicle, equipped with a reusable first stage.
The mission involves injecting a 6-ton payload into an 800~km circular sun-synchronous orbit, with launch operations taking place from the European spaceport in French Guiana.
The recovery of the first stage is achieved through a glide-back \citep{baioccoOverview2021}.

\begin{figure}[!ht]
    \centering
    \begin{tabular}[t]{cc}
            \includegraphics[width=0.33\linewidth,valign=t]{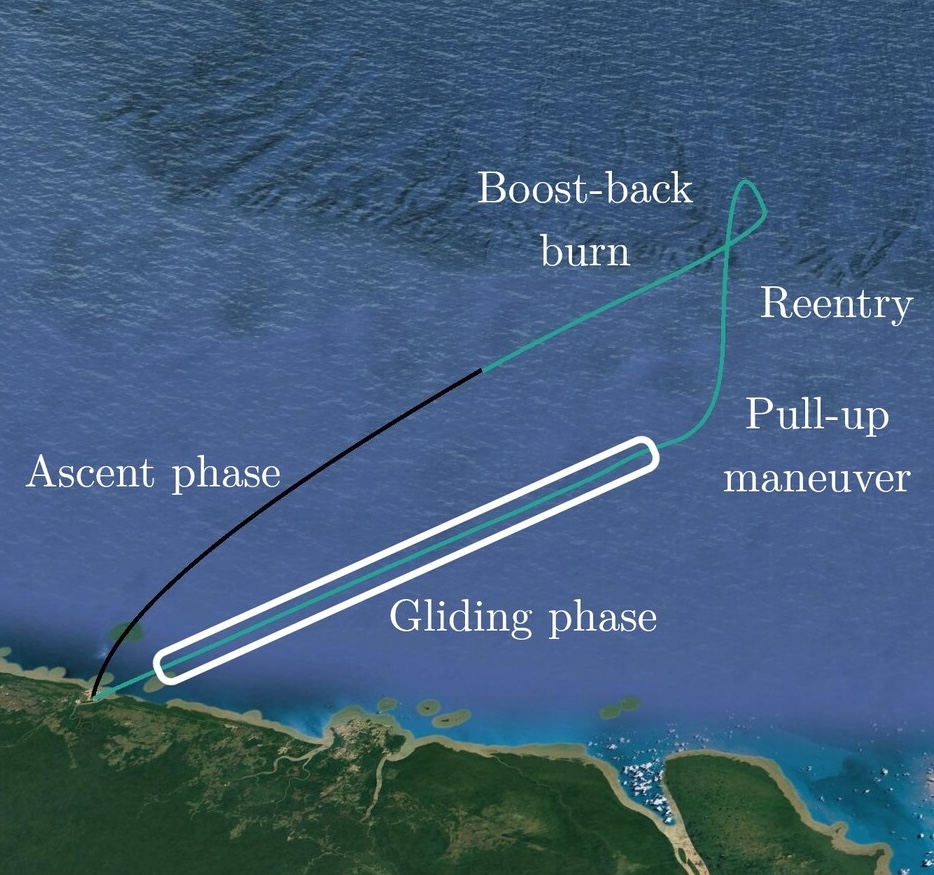} &
            \includegraphics[width=0.35\linewidth,valign=t]{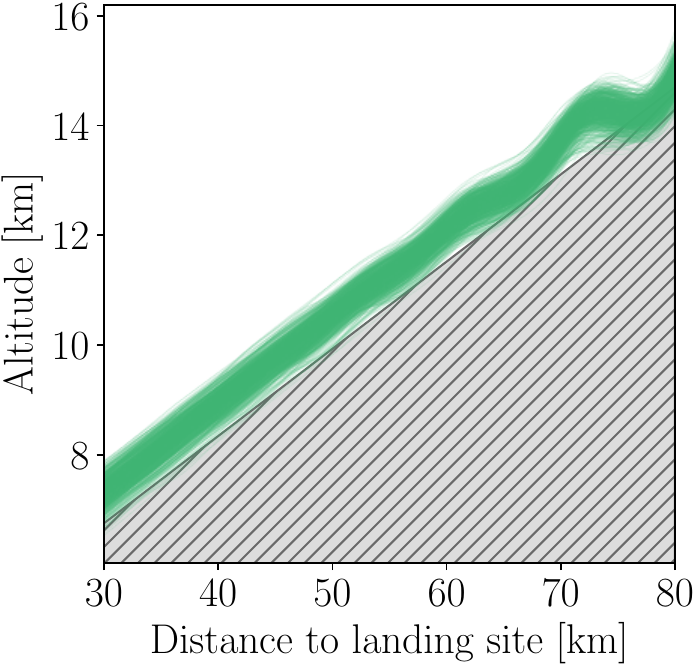}
    \end{tabular}
    \caption{Different flight phases (left) and samples of the launcher trajectory with the flight corridor (right).}
    \label{fig:launcher}
\end{figure}

After separation from the second stage, the first stage turns around and performs a boost-back maneuver to reverse its velocity and move toward the landing site.
This is followed by an atmospheric re-entry, a pull-up maneuver, and a glide phase until reaching the landing site.
A final alignment phase prepares the vehicle for runway landing.
An illustration of the different flight phases is provided on the left of \Cref{fig:launcher}.
The complete trajectory is computed by numerically solving the three degrees of freedom equations of motion using a fifth-order variable-step Runge-Kutta integrator \citep{atkinson2008introduction}.
The underlying physical model is detailed in \citet{balesdentMultidisciplinary2023}.
This study specifically focuses on the glide-back segment of the trajectory.
For this purpose, only the portion of the altitude at a distance between 80 km and 30 km from the landing site is extracted.
To ensure a consistent spatial discretization across different simulations, all trajectories are linearly interpolated onto a uniform grid of 1,001 nodes.

The uncertain input vector $\bm U$ includes 14 variables: the aerodynamic axial-force coefficient $C_a\sim \mathcal N(1, 0.02^2)$, the aerodynamic normal-force coefficient $C_n\sim \mathcal N(1.1, 0.02^2)$, the flight conditions detailed right after, the thrust vector angle $\omega\sim \mathcal N(0,0.5^2)$ (in degrees), and the mass of unburnt fuel $M_\mathrm{unburnt}\sim \mathcal N(0, 10^2)$ (in kg), truncated at 0.
Uncertainty in flight conditions is modeled via perturbations in the angle of attack (which can be assimilated to wind perturbations), represented as a linear interpolation of ten independent control points, whose values are denoted by $\delta_i\sim \mathcal N(1, 0.5^2)$, $i\in\{1,\dots,10\}$.
The complete 14-dimensional input vector is therefore $\bm U=[C_a, C_n, \delta_1,\dots,\delta_{10},\omega,M_\mathrm{unburnt}]^\top$.

A flight corridor is established for population and facility safety, as well as air traffic control purposes.
Note that the definition of this corridor does not correspond to any real mission, yet it is considered realistic.
It begins at an altitude of 6.75 km, located 30 km away from the launch and landing site, and rises to 14.7 km at a distance of 80 km.

Regarding active learning, the initial DoE consists of $\dim(\mathbb U)=14$ samples.
The strategy adds $10\times\dim(\mathbb U)=90$ additional samples.
The Monte Carlo sample set size is set to $n_\mathrm{MCS}=40{,}000$.
The Kriging kernel function is Mat\'ern-5/2 and the nugget factor is set equal to $10^{-10}$.
PCA is performed with a RIC of 99.99\%, which for 100 samples results in 15 principal directions.

\begin{figure}[!ht]
    \centering
    \includegraphics[height=5.2cm]{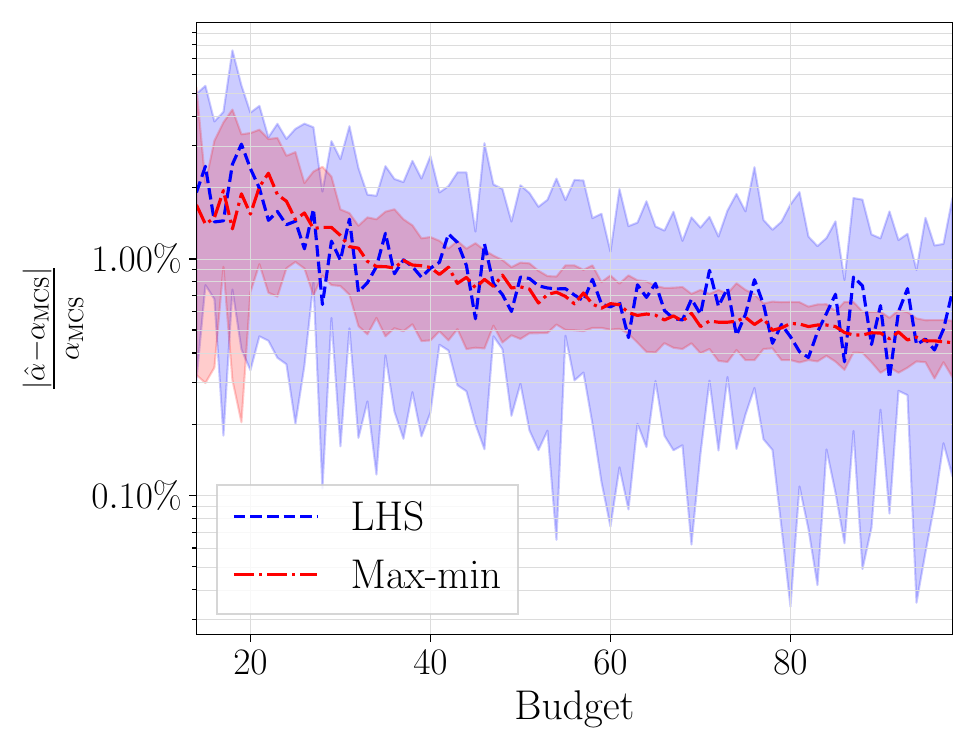}
    \hspace{1cm}
    \includegraphics[height=5.2cm]{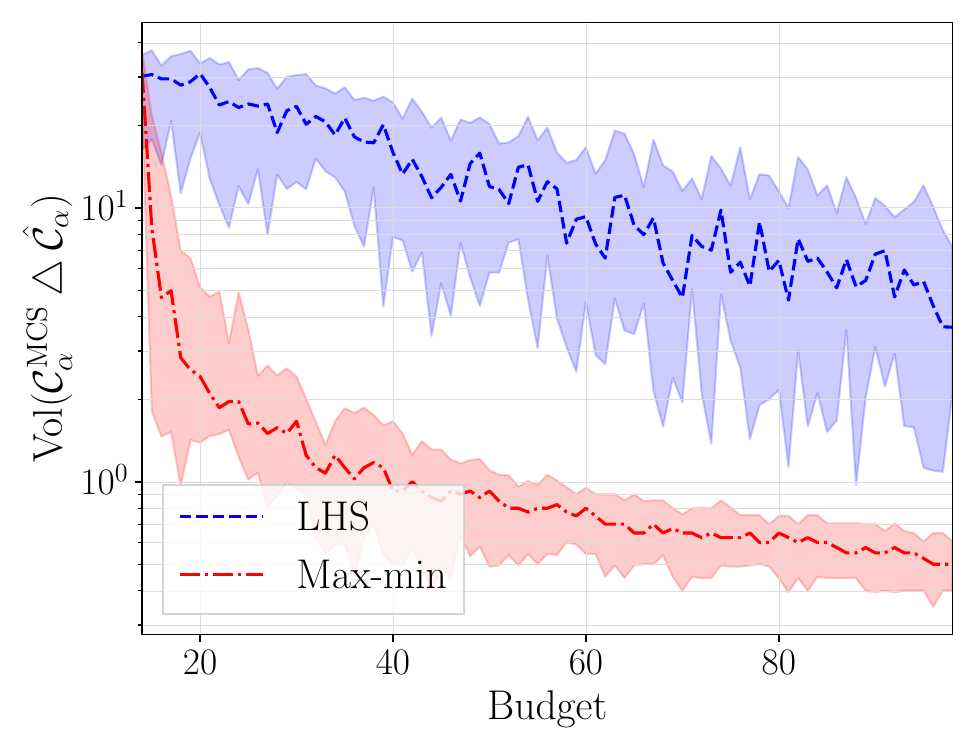}
    \caption{Convergence of $\hat{\alpha}$ (left) and $\Vol(\mathcal{C}_\alpha\bigtriangleup \hat{\mathcal{C}}_\alpha)$ (right, in km) for the launcher case study. The line is the median across the repetitions and the area between the 0.1- and 0.9-quantile is shaded.}
    \label{fig:launcher-a-symdiff}
\end{figure}

First, the convergence of $\hat{\alpha}$ is shown on the left of \Cref{fig:launcher-a-symdiff} in logarithmic scale.
Starting from 14 training samples, the initial error is around 2\% for the LHS (blue) and max-min (red) strategies as they have the same initial training set.
Throughout the entire learning process, both methods yield a similar level of median relative error, indicating that both have a comparable error in the probability of containing the true excursion set.
After the addition of 80 training points, the relative error in $\alpha$ is around 0.5\% for both approaches.
However, our approach shows a significantly better robustness across repetitions, as can be seen in \Cref{fig:launcher-a-symdiff} (left) from the much smaller 0.1- to 0.9-quantile range in red.
The analysis of the volume of the symmetric difference between the reference confidence region and its estimate, shown on the right of \Cref{fig:launcher-a-symdiff}, reveals a much more significant final accuracy of our active learning approach.

The evolution of $\Vol(\mathcal{C}_\alpha\bigtriangleup \hat{\mathcal{C}}_\alpha)$ (in km) between the reference confidence region and its estimate is shown on the right of \Cref{fig:launcher-a-symdiff}, also in logarithmic scale.
The two strategies start with a volume of approximately 30~km.
Since the trajectory of the launcher is considered between 30 and 80~km, this corresponds to a significant 60\% of the entire mesh.
As mentioned above, the convergence of this metric differs from that of $\hat{\alpha}$.
The difference between our method (red) and the naive LHS strategy (blue) is clear from the first few iterations.
With the naive LHS, the volume decreases slowly to around 4~km, whereas in our approach it reduces to approximately 0.5~km by the end of the iterations, which is more likely to be compliant with safety regulations.
This corresponds to an error of about 8\% of the mesh volume for the former, compared to only 1\% for our method.

\begin{figure}[!ht]
    \centering
    \includegraphics[height=5.2cm]{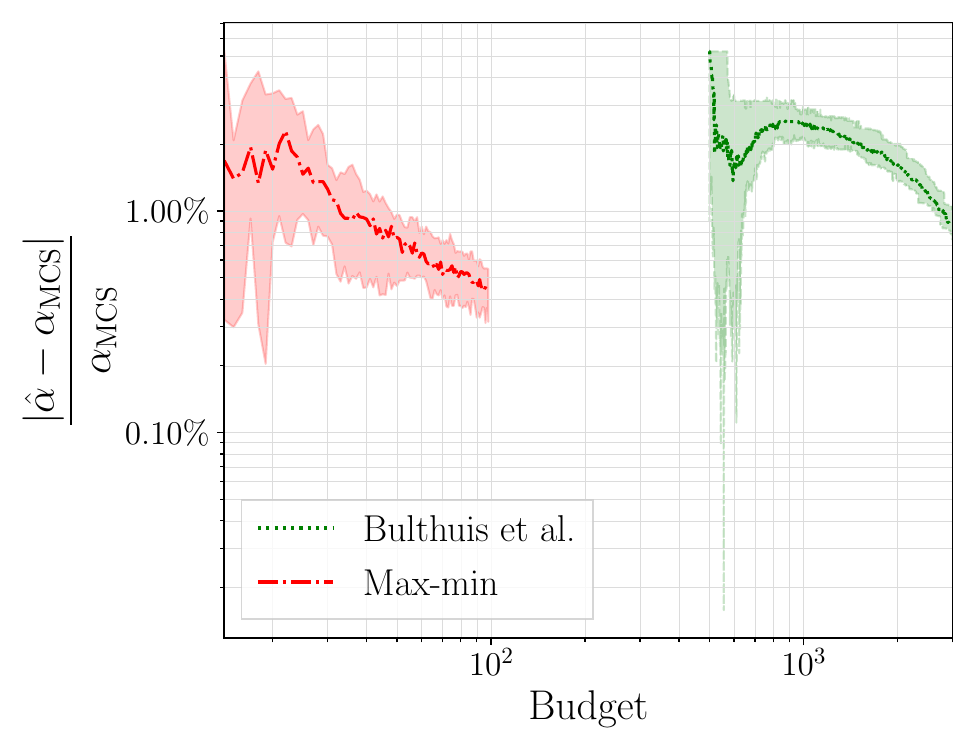}
    \hspace{1cm}
    \includegraphics[height=5.2cm]{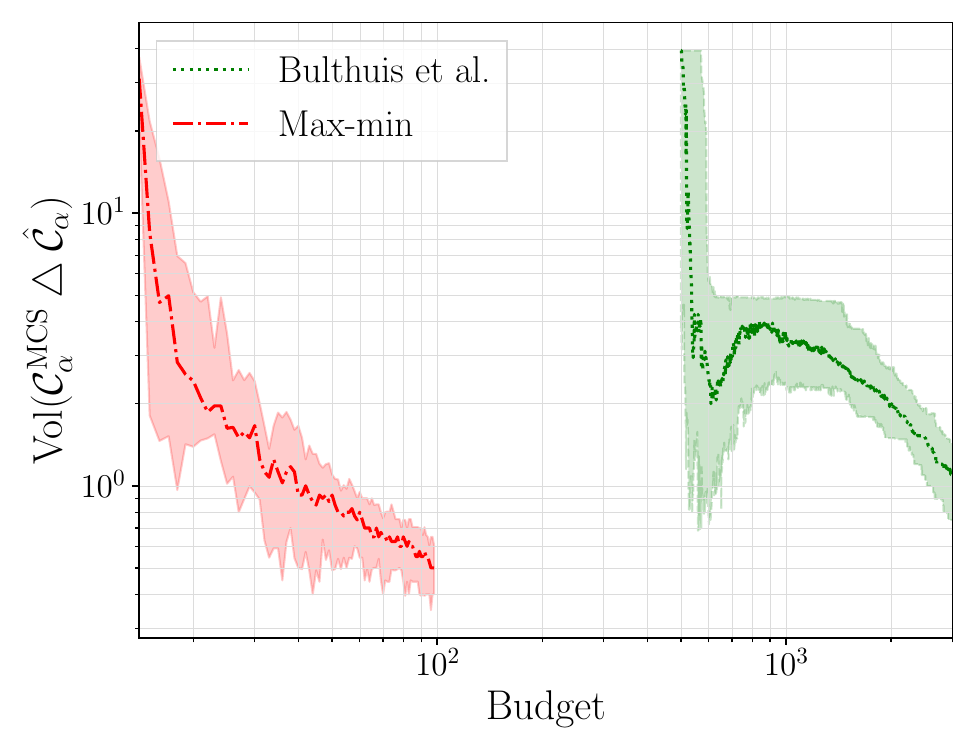}
    \caption{Convergence of $\hat{\alpha}$ (left) and $\Vol(\mathcal{C}_\alpha\bigtriangleup \hat{\mathcal{C}}_\alpha)$ (right) for the launcher case study, for our method in red, and the one proposed in \citet{bulthuisMultifidelity2020} in green. The line is the median across the repetitions and the area between the 0.1- and 0.9-quantile is shaded.}
    \label{fig:launcher-bulthuis}
\end{figure}

We compare our results with those obtained using the method of \citet{bulthuisMultifidelity2020} in \Cref{fig:launcher-bulthuis}.
The convergence of $\hat{\alpha}$ is shown on the left, and $\Vol(\mathcal{C}_\alpha\bigtriangleup \hat{\mathcal{C}}_\alpha)$ is shown on the right.
First, let us analyze $\hat{\alpha}$.
The approach of \citet{bulthuisMultifidelity2020} starts with a median (across 20 repetitions) relative error around 5\% with 500 training samples, and decreases to approximately 0.9\% with 3{,}000 additional points, amounting to a total of 3{,}500.
In comparison, our method achieves a better level of accuracy for about 36 times less points, terminating with a median relative error around 0.5\% with a total of 94 training samples.

The analysis of the symmetric difference, shown on the right of \Cref{fig:launcher-bulthuis}, draws a similar picture.
The algorithm of \citet{bulthuisMultifidelity2020} starts with approximately 40~km for 500 training points, which decreases to just above 1~km after the addition of 3{,}000 more samples.
In contrast, using only 94 training points, our method has a smaller median symmetric difference of 0.5~km.
Overall, both metrics indicate a greater efficiency of our proposed methodology.

\begin{figure}[p]
    \centering
    \includegraphics[width=\textwidth]{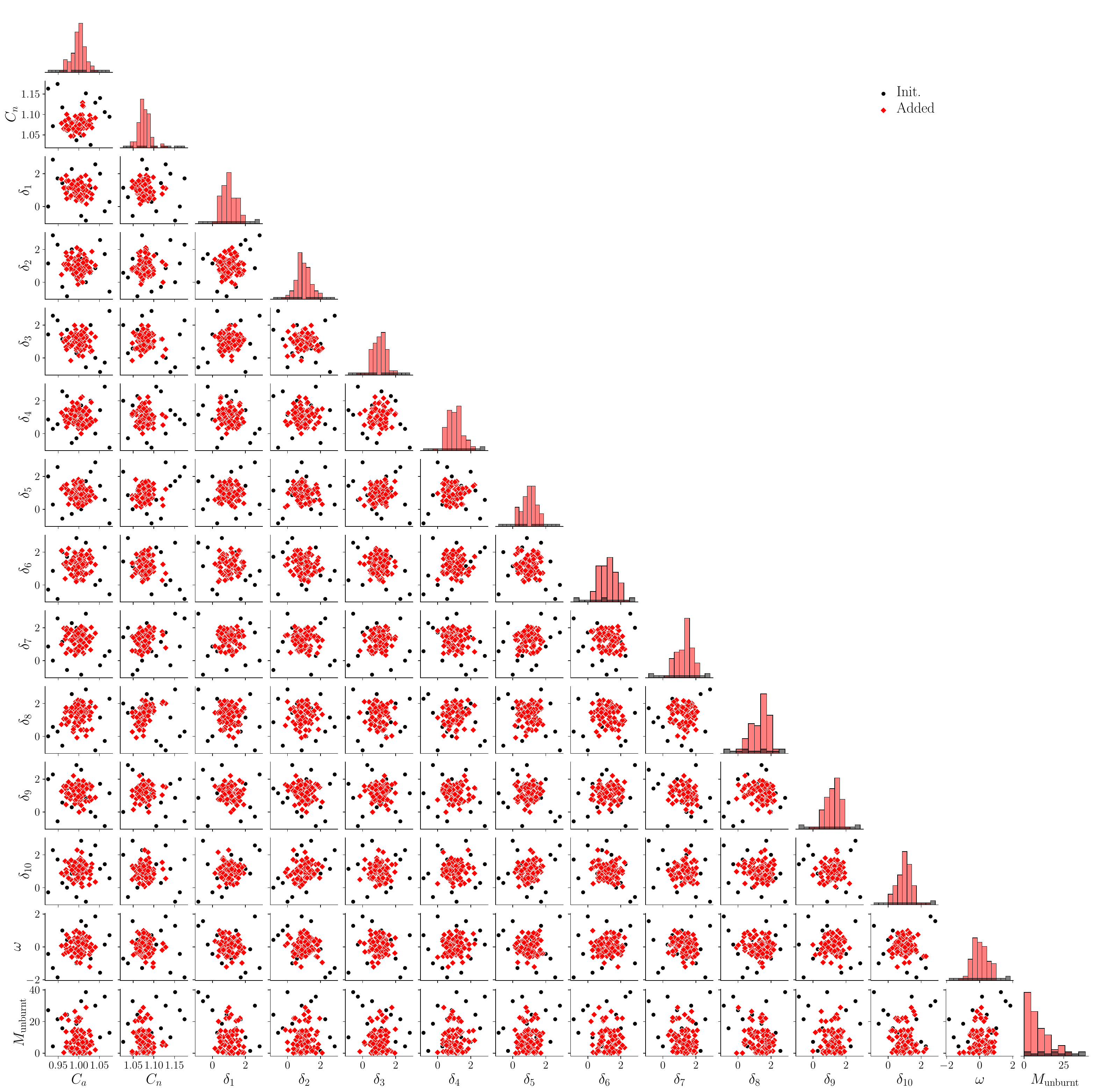}
    \caption{Initial and added samples locations in $\mathbb{U}$ for the launcher case.}
    \label{fig:pairplot-excursion-launcher}
\end{figure}

In \Cref{fig:pairplot-excursion-launcher}, we show where the training samples are located in the input domain~$\mathbb{U}$ for our strategy only, as a similar graph for the LHS approach does not provide much information.
The max-min strategy leads to the selection of low normal-force coefficients $C_n$ and mass of unburnt fuel $M_\mathrm{unburnt}$.
In addition, it appears that the wind perturbations $\delta_7$ to $\delta_{10}$ -- corresponding to roughly the last third of the trajectory in terms of distance to launch site -- have a significant impact on the confidence region, as suggested by the skewed bar charts on the diagonal of \Cref{fig:pairplot-excursion-launcher}.
Other variables tend to be selected with a distribution similar to the input one.

\begin{figure}[!ht]
    \centering
    \begin{tabular}{rr}
        \includegraphics[width=0.335\linewidth]{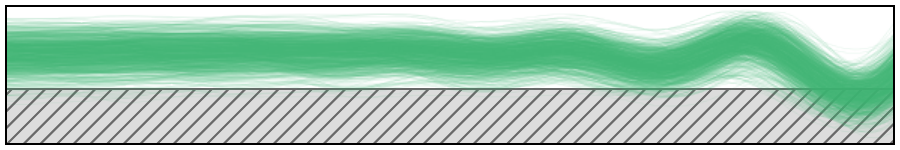}\; & \includegraphics[width=0.335\linewidth]{_launcher_snap_low.pdf}\; \\
        \includegraphics[width=0.35\linewidth]{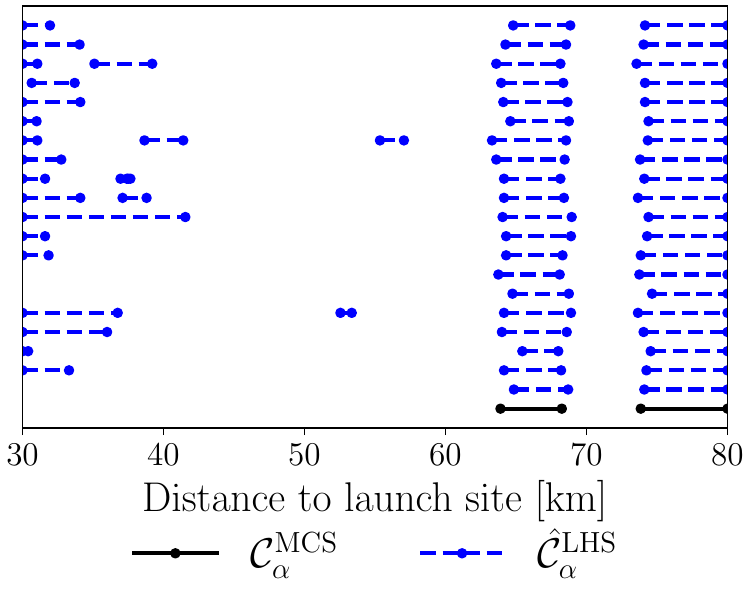} & \includegraphics[width=0.35\linewidth]{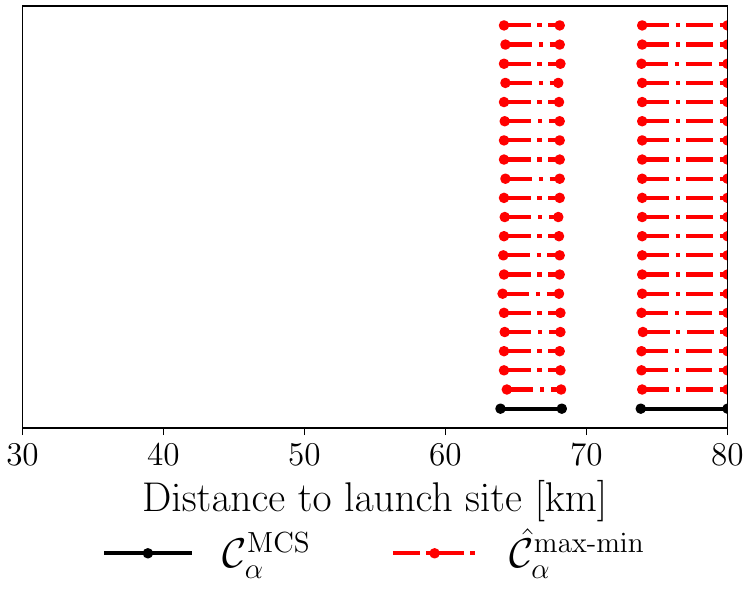} \\
        \\
        \includegraphics[width=0.412\linewidth]{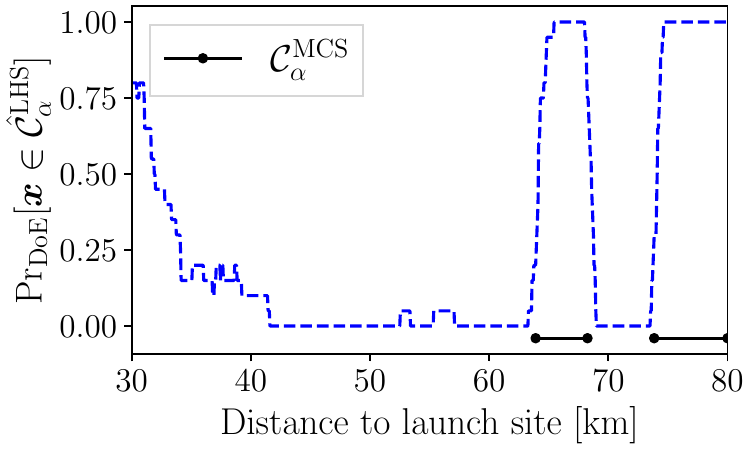} & \includegraphics[width=0.412\linewidth]{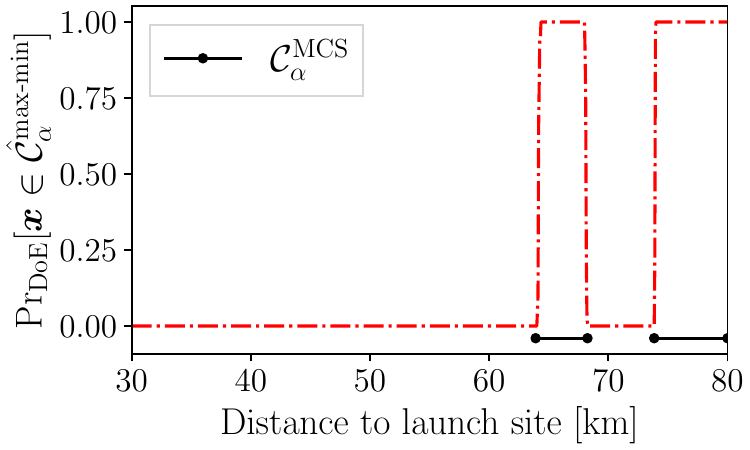}
    \end{tabular}
    \caption{Variability of the estimated confidence region to the choice of the initial DoE. Top: boundaries of the confidence region obtained with the naive LHS (left) and our active learning strategy (right) across 20 repetitions. Bottom: probability that each mesh node belongs to the estimated confidence region, computed from the 20 repetitions using the naive LHS (left) and our approach (right).}
    \label{fig:launcher-C-doe}
\end{figure}

As in the previous case studies, the final confidence regions corresponding to the 20~different initial DoEs are shown in the top row of \Cref{fig:launcher-C-doe}.
On the left and in blue, the confidence regions generated using the naive LHS approach exhibit noticeably greater variability than those obtained with our strategy, shown on the right in red.
Beyond the reduced variability, the confidence regions produced by our method also appear more accurate.
The bottom row of \Cref{fig:launcher-C-doe} depicts the probability that each mesh node belongs to the confidence region, accounting for the uncertainty introduced by the choice of initial DoE.
These probabilities are computed from the 20 repetitions shown above.
For the naive LHS approach, the probability is far from 0 or 1 between 30 and 42~km, explaining the high variability observed in the corresponding confidence regions above.
In contrast, our strategy yields probabilities that are almost exclusively close to 0 or 1, suggesting strong confidence in the topology of the estimated confidence region.

\Cref{fig:launcher-C-doe} also shows how the median $\hat{\alpha}$ across repetitions can appear accurate even when the symmetric difference is large.
Most of the LHS-based confidence regions include elements between 30 and 60~km, whereas the reference region does not.
Because the right part of the confidence region is somewhat not accurate, it fails to include some excursion sets that it should contain.
At the same time, some excursions sets appearing at smaller distances to launch site are included even though they should not be.
Overall, these opposite effects tend to compensate for each other, resulting in an accurate $\hat{\alpha}$, although the geometry of the confidence region noticeably differs from the reference.

\begin{figure}[!ht]
    \centering
    \begin{tabular}{rr}
        \includegraphics[width=0.335\linewidth]{_launcher_snap_low.pdf}\; & \includegraphics[width=0.335\linewidth]{_launcher_snap_low.pdf}\; \\
        \includegraphics[width=0.35\linewidth]{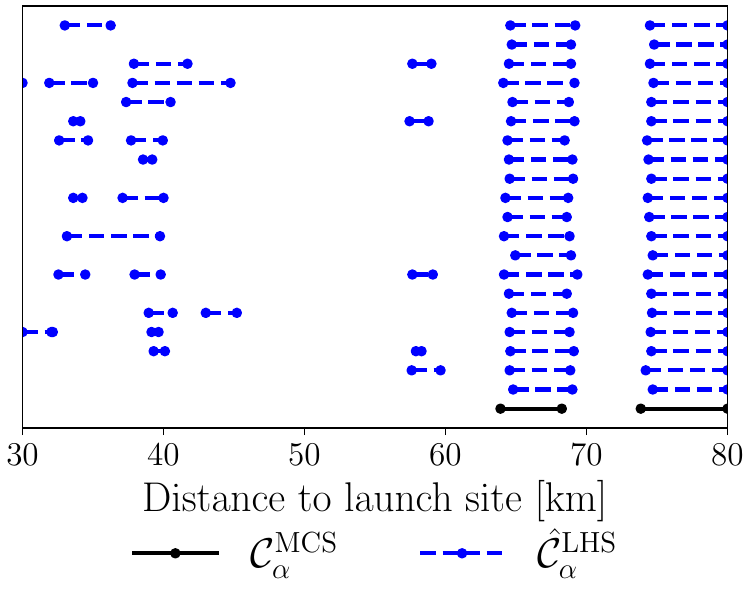} & \includegraphics[width=0.35\linewidth]{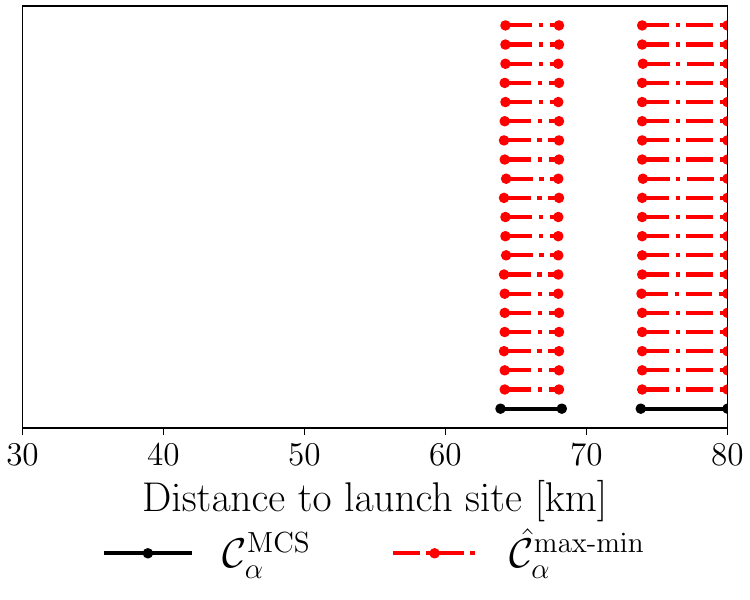} \\
        \\
        \includegraphics[width=0.412\linewidth]{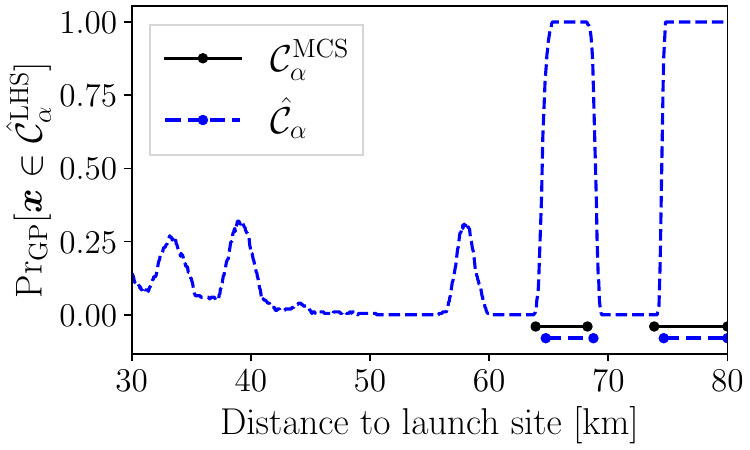} & \includegraphics[width=0.412\linewidth]{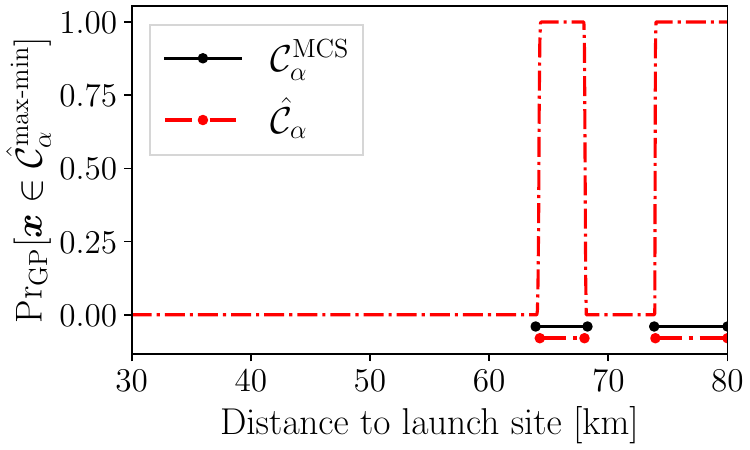}
    \end{tabular}
    \caption{Variability of the estimated confidence region to the intrinsic variability of the GPs for one particular initial DoE. Top: boundaries of the confidence region obtained with the naive LHS (left) and our active learning strategy (right) across 200 realizations of the GPs. Bottom: probability that each mesh node belongs to the estimated confidence region, computed from 200 realizations using the naive LHS (left) and our approach (right).}
    \label{fig:launcher-C-gp}
\end{figure}

Finally, the impact of the intrinsic uncertainty of the GPs is illustrated in \Cref{fig:launcher-C-gp}.
The top row displays the confidence regions corresponding to 20 realizations of the GPs, trained with the median DoE.
As with the previous metric, the naive LHS strategy (blue, on the left) results in significant variability, particularly between 30 and 45~km and around 58~km.
In contrast, the confidence regions generated by our method (red, on the right) exhibit very little variability.
The bottom row of \Cref{fig:launcher-C-gp} shows the probability of each mesh node belonging to the confidence region approximated using 200 GP realizations, for both the naive LHS approach (left) and our strategy (right).
For the naive LHS, the conclusion remains: the variability due to the GPs is noticeable in the 30-45~km range and around 58~km.
As a result, combining the areas where the confidence region is and where it might be, it spans a significant portion of the entire mesh.
On the other hand, our approach produces a stable confidence region across most of the domain, with probabilities everywhere close to either 0 or 1.

\section{Conclusion and outlook}\label{sec:conclusion}

This paper proposed a surrogate-based approach for estimating excursion-set confidence regions in scenarios where the simulator output is functional and must be evaluated over the entire mesh at once.
The surrogate used to replace the costly simulator combines principal components analysis with Kriging.
An active learning strategy was introduced, leveraging efficient sampling of Gaussian processes realizations via a Karhunen-Lo\`eve expansion with Nystr\"om approximation, propagated through all the quantities up to the confidence region.
The strategy selects points aiming to reduce the predictive uncertainty of $\hat\rho^*$, a quantile of a random variable linked to the excursion set required to compute the confidence region, while ensuring diversity with its max-min nature.

The approach was applied to three case studies: one analytical, and two industrial ones, namely, the pressure coefficient distribution over a hypersonic vehicle and the return-to-launch-site trajectory of a reusable launcher first stage.
In all cases considered, the strategy demonstrated better performance than a reference naive approach, and outperformed by one or two orders of magnitude in efficiency the only applicable method of the literature.
Relevant metrics were discussed to assess its effectiveness: the effective containment probability, the volume of the symmetric difference between the reference region and its estimate, and the remaining variability caused by the choice of initial DoE and the GP surrogate modeling.
The space launcher case serves to illustrate the importance of these various metrics.

A current limitation of this work is the reliance on crude Monte Carlo simulation.
Indeed, achieving a reasonably accurate estimate of the confidence region requires a large number of samples and many corresponding realizations of the Gaussian processes.
This poses a challenge when dealing with very rare events.
It is also problematic regarding memory requirements: the Monte Carlo sample matrix in the hypersonic case requires almost 18~GB of storage space or memory.
Similarly, the numerical sampling approach used to propagate the GP predictive uncertainty could be improved by developing an analytical solution.
Another limitation is that the surrogate modeling uncertainty is restricted to the propagated GP predictive uncertainty, ignoring other sources (mostly the Monte Carlo size, the dimensionality reduction and the parametrization of the confidence region).
Even though the linear and unsupervised PCA was accurate enough for our numerical experiments, it might be insufficient for certain functional outputs, notably those with features moving along the mesh.
Furthermore, the scalability of the proposed approach with respect to the input dimensionality is constrained by the latent GPs. Traditional Kriging models are known to scale poorly in high-dimensional input spaces.
Finally, the method only uses a maximum budget as a stopping criterion.
As we show in the numerical experiments, choosing a maximum budget is difficult for confidence regions as different metrics give different insights into the learning process.
These topics are all candidates for future research.

\section*{Acknowledgements}

This work is co-funded by ONERA and the Agence de l'innovation de
défense.

\begin{appendices}
\crefalias{section}{appendix}

\section{Definition of the auxiliary random variable}\label{app:proof}

    In \Cref{sec:surrogate-estimation}, to be able to work with empty excursion sets, we have introduced the following definition of the auxiliary variable:
    \begin{equation}
        \begin{split}
            \hat{\chi}:{} & \mathbb{U}\rightarrow[0,1] \\
            & \bm{U}\mapsto\left\{
                \begin{matrix*}[l]
                    \min\{\hat p(\bm x):\bm x\in\hat{\mathcal E}(\bm U)\} & \text{if $\hat{\mathcal{E}}(\bm{U})\neq\emptyset$},\\
                    1 & \text{otherwise.}
                \end{matrix*}
            \right.
        \end{split}
    \end{equation}

    To reformulate the confidence region estimation as a quantile problem using \Cref{eq:simplification}, we first need to prove that we can rewrite $\Pr[\hat{\mathcal{E}}(\bm{U})\subset\mathcal{R}]$ as $1-\Pr[\hat{\chi}(\bm{U})<\rho]$, where $\rho$ is the Vorob'ev threshold.
    To this aim, we need to show that the condition $[\hat{\mathcal{E}}(\bm{U})\subset\mathcal{R}]$ is equivalent to $[\hat{\chi}(\bm{U})\geq\rho]$.
    To make the calculations clearer, we introduce the event $\omega\in\Omega$ from probability theory in the notation.
    We can write:
    \begin{equation}
        \mathcal{G}\stackrel{\mathrm{def}}{=}
        [\hat{\mathcal{E}}(\bm{U})\subset\mathcal{R}]=\{\omega\in\Omega:\forall\bm{x}\in\hat{\mathcal{E}}(\bm{U}(\omega)):\hat{p}(\bm{x})\geq\rho\},
    \end{equation}
    and
    \begin{equation}
        \mathcal{H}\stackrel{\mathrm{def}}{=}[\hat{\chi}(\bm{U})\geq\rho]=\{\omega\in\Omega:\hat{\chi}(\bm{U}(\omega))\geq\rho\}.
    \end{equation}

    First, let us prove that $\mathcal{G\subset H}$.
    By definition, if $\omega\in\mathcal{G}$, then $\forall\bm{x}\in\hat{\mathcal{E}}(\bm{U}(\omega))$ we have $\hat{p}(\bm{x})\geq\rho$.
    If $\hat{\mathcal{E}}(\bm{U}(\omega))=\emptyset$, by \Cref{eq:def-chi} we have $\hat{\chi}(\bm{U}(\omega))=1$.
    As $\rho\in[0,1]$, then $\hat{\chi}(\bm{U}(\omega))\geq \rho$.
    If $\hat{\mathcal{E}}(\bm{U}(\omega))\neq\emptyset$, then $\hat{\chi}(\bm{U}(\omega))=\min\{\hat{p}(\bm{x}):\bm{x}\in\mathcal{E}(\bm{U}(\omega))\}$.
    By the definition of $\mathcal{G}$, $\forall\bm{x}\in\hat{\mathcal{E}}(\bm{U}(\omega))$ we have $\hat{p}(\bm{x})\geq\rho$, so $\hat{\chi}(\bm{U}(\omega))\geq\rho$.
    We can see that whether $\hat{\mathcal{E}}(\bm{U}(\omega))$ is empty or not, $\omega$ is also in $\mathcal{H}$.
    Consequently, $\mathcal{G\subset H}$.

    Then, let us prove that $\mathcal{H\subset G}$.
    By definition, $\omega\in\mathcal{H}$ implies that $\hat{\chi}(\bm{U}(\omega))\geq\rho$.
    Since $[\forall\bm{x}\in\emptyset,S(\bm{x})]$ is true for any statement $S(\bm{x})$ (this is called a \emph{vacuous truth}; it is true because one cannot prove its contrapositive that there exists at least one $\bm{x}$ for which $S(\bm{x})$ is not true), then we have $[\forall\bm{x}\in\mathcal{E}(\bm{U}(\omega)):\hat{p}(\bm{x})\geq\rho]$.
    If $\mathcal{E}(\bm{U}(\omega))\neq\emptyset$, by the definition of $\mathcal{H}$ we know that $\hat{\chi}(\bm{U}(\omega))\geq\rho$.
    Since $\hat{p}(\bm{x})\geq\hat{\chi}(\bm{U}(\omega))$, then $\hat{p}(\bm{x})\geq\rho$.
    Whether $\hat{\mathcal{E}}(\bm{U}(\omega))$ is empty or not, $\omega$ is also in $\mathcal{G}$.
    Consequently, $\mathcal{H\subset G}$.

    To conclude, since $\mathcal{G\subset H}$ and $\mathcal{H\subset G}$, then $\mathcal{G=H}$.
    We can thus rewrite $\Pr[\hat{\mathcal{E}}(\bm{U})\subset\mathcal{R}]$ as $\Pr[\hat{\chi}(\bm{U})\geq\rho]$, or $1-\Pr[\hat{\chi}(\bm{U})<\rho]$.

\section{Comparative analysis of the auxiliary variable field}\label{app:chi}

\begin{figure}[!ht]
        \centering
        \includegraphics[width=0.9\textwidth]{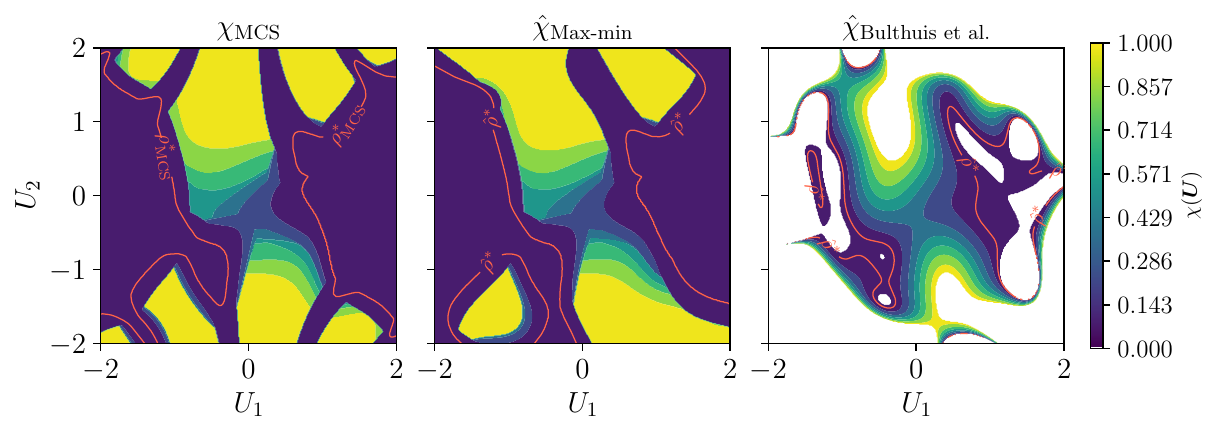}
        \caption{Map of $\chi$ across the initial DoE domain with the Monte Carlo estimate using the true simulator on the left, estimated using our approach in the center (using sampling for optimization), and using PCE model of $\chi$ in the method of \citet{bulthuisMultifidelity2020} on the right (values below 0 or above 1 are excluded).}
        \label{fig:normsfast-chi}
    \end{figure}

    To help better understand the difference in performance between our method and that of \citet{bulthuisMultifidelity2020}, we show the map of $\chi$ within $[-2,2]^2$ in \Cref{fig:normsfast-chi} for the first analytical example.
    The Monte Carlo-based reference is on the left, while the estimates using our method and the one of \citet{bulthuisMultifidelity2020} are on the center and the right, respectively.
    Looking at the reference values of $\chi_\mathrm{MCS}(\bm{U})$, we can see that $\chi$ is a piecewise continuous function of the input.
    Additionally, according to the definition of this auxiliary variable, it takes values within $[0,1]$.
    The estimate with our method shown in the middle is very accurate around the center of the domain but less so near the boundaries.
    This is expected since samples are added closer to the center, where the input probability density is highest, while outer regions where little to no Monte Carlo samples are located are mostly left unexplored.

    On the right, we display the PCE estimate of the auxiliary variable, with values above 1 or below 0 ignored to improve readability.
    Near the center of the domain, the estimated level set is similar to the true one though not close, showing a poorer approximation of $\chi$.
    Their algorithm converges anyway for a large number of iterations because the PCE-based values of $\hat{\chi}$ are replaced by the value obtained using the true simulator.
    Besides, the use of a surrogate to model $\chi$ leads to values below 0 or above 1, which should not be possible given the definition of $\chi$ in \Cref{eq:def-chi}.

\end{appendices}

\bibliographystyle{abbrvnat}
\bibliography{biblio}

\end{document}